\documentclass[%
reprint,
amsmath,amssymb,
aps,
prb, two-column,
dvipsnames,
floatfix,
]{revtex4-2}

\usepackage{graphicx}
\usepackage{bm}
\usepackage{hyperref}

\usepackage[dvipsnames]{xcolor}
\hypersetup{
	colorlinks,
	linkcolor={red!80!black},
	urlcolor={blue!80!black},
	citecolor= {blue!80!black} 
}

\begin{document}
\title{Nonlinear Hall effect in isotropic k-cubed Rashba model: Berry-curvature-dipole engineering by in-plane magnetic field}

\author{A. Krzyżewska and A. Dyrdał}
\email[corresponding author: ]{anna.dyrdal@amu.edu.pl}
\affiliation{ 
Faculty of Physics, ISQI,
Adam Mickiewicz University in Poznań, ul. Uniwersytetu Pozna\'nskiego 2, 61-614
Pozna\'n, Poland
}

\begin{abstract}
The linear and nonlinear Hall effects in 2D systems are considered theoretically within the isotropic k-cubed Rashba model. We show that the presence of an out-of-plane external magnetic field or net magnetization is a necessary condition to induce a nonzero Berry curvature in the system, whereas an in-plane magnetic field tunes the Berry curvature leading to the Berry curvature dipole. 
Interestingly, in the linear response regime, the conductivity is dominated by the intrinsic component (Berry curvature component), whereas the second-order correction to the Hall current (i.e., the conductivity proportional to the external electric field) is dominated by the component independent of the  Berry curvature dipole.
\end{abstract}
\maketitle

\section{Introduction}
\label{sec:introduction}

Spin-orbit interaction of Rashba type~\cite{Bychkov1984,Winkler2003,Bihlmayer2015May} is a consequence of structural inversion asymmetry in a system and  originally has been studied in the context of semiconductor surfaces and heterostructures with asymmetric confining potentials. In the current relevant literature, the Rashba spin-orbit coupling means usually the spin-orbital interaction arising in systems without the inversion centre, that can be described as an effective internal and odd-momentum-dependent Zeeman-like field acting on the electron spin~\cite{Manchon_NatMat2015,Bihlmayer_NatRevPhys2022}. Consequently, the spin-orbit interaction leads the  so-called spin-momentum locking phenomenon, i.e., to a fixed orientation of quasiparticle spin with respect to its momentum. This, in turn, is responsible for fascinating electrical and optical effects, that are currently a hallmark of spin-orbitronics~\cite{Soumyanarayanan2016Nov,Feng2017Sep,Hirohata2020Sep,Hsieh2009Aug,Bernevig2005Jul,Chernyshov2009Sep}. In the context of electric transport, especially important consequences of spin-momentum locking are the spin-to-charge interconversion effects, such as current-induced spin polarization (also known as Edelstein effect)\cite{Kato2004Oct,Sih2005Oct,Li2014Mar} and spin Hall effect~\cite{Sinova_RevModPhys_SHE_2015,Kato2004Dec,Wunderlich2005Feb,Sih2005Oct,Valenzuela2006Jul,Saitoh2006May,Wunderlich2010Dec}. 
Recently, in the context of spin-orbitronics, nonlinear transport effects, i.e., bilinear magnetotransport~\cite{He2018May,Dyrdal2020Jan,Fu2022Oct,Guillet2020Jan,Vaz2020Jul}, and nonlinear Hall effect are of special attention. 

The nonlinear Hall effect~\cite{Sodemann2015Nov,Ma2019Jan,Kang2019Apr,He2019Jul,Du2021Aug} (NLHE) is an intriguing member of the family of Hall effects, as it can arise in time-reversal symmetric systems. The only condition that is required is the absence of inversion symmetry~\cite{Sodemann2015Nov, Ma2019Jan}. 
Similarly to the anomalous Hall effect~\cite{Sinova_RevModPhys_AHE_2010} (AHE) one can define intrinsic and extrinsic microscopic mechanisms responsible for NLHE~\cite{Du2019Jul,Du2021Aug}. The intrinsic contribution has a geometric nature, namely, the positive and negative Berry curvature hotspots are located in slightly different regions of the Brillouin zone, leading to the dipole moment that is called Berry curvature dipole (BCD)~\cite{Sodemann2015Nov}. This intrinsic contribution, even originating in case of nonzero Berry curvature, is different from the intrinsic AHE, where the conductivity is robust to scattering and is dissipationless. This is because the intrinsic component of NLHE is nonzero only when the Fermi level crosses the energy bands (i.e., the BCD disappears in the energy gap), and consequently, the effects of disorder are unavoidable. The BCD has been reported in topological crystalline insulators (e.g., the (001) surface of SnTe, Pb$_{1-x}$Sn$_x$Te, Pb$_{1-x}$Sn$_x$Se)~\cite{Sodemann2015Nov,Du2019Jul}, Weyl semimetals~\cite{Zhang2018Jan}, and transition-metal dichalcogenides~\cite{You2018Sep,Zhou2020Feb,Kang2019Apr,Nandy2019Nov,Du2019Jul}.
In turn, the extrinsic contribution to NLHE is related to spin-dependent scattering processes, also well-known from the theory of AHE, i.e., to the {\textit{skew-scattering}} and {\textit{side jump}} ~\cite{Nandy2019Nov,Kang2019Apr,He2019Jul,Du2019Jul}. 

In this work, we develop the concept of externally controlled nonlinear Hall effect. We show, that the Berry curvature and Berry curvature dipole can emerge as a consequence of external force, such as external magnetic field, which additionally controls the properties of BCD and nonlinear system response. Such external control of nonlinear (unidirectional) electronic transport provides an additional degree of freedom in the design of new spintronics devices.  
Accordingly, we demonstrate the Berry curvature dipole engineering by external magnetic field and present a detailed study of the linear and nonlinear Hall response in two-dimensional electron gas with isotropic cubic form of Rashba spin-orbit interaction. The Berry curvature in such a system appears as a consequence of nonzero out-of-plane external magnetic field or magnetization. In turn, the Berry curvature dipole emerges (and simultaneously can be controlled) due to an in-plane magnetic field. In consequence, the Hall conductivity contains linear and nonlinear components. The linear system response contains the contributions due to  anomalous and planar Hall effects, whereas the nonlinear Hall response is determined by BCD. 

Our proposition of magnetic control of Berry curvature and Berry curvature dipole is quite general and can be applied to any kind of systems that, under zero external fields, reveal only the lack of  spatial inversion symmetry.  Thus we decided to consider here the effective model describing 2D quasiparticles in the presence of isotropic cubic form of Rashba SOC. This kind of Rashba spin-orbit coupling determines the spectra of p-states in p-doped zincblende III-V semiconductor heterostructures  (heavy holes in 2DHG)~\cite{Winkler2003,Bernevig2006Dec,Schliemann2005Feb,Liu2008Mar,Moriya2014Aug}. Interestingly, the same kind of spin-orbital splitting, i.e., the same symmetry of low-energy electronic states, associated with $t_{2g}$ orbitals, is observed for 2D electron gas emerging at the surface states of cubic perovskites, such as ${\mathrm{SrTiO_{3}}}$ (STO) and STO-based interfaces, e.g., ${\mathrm{LaAlO_{3}/SrTiO_{3}}}$~\cite{vanHeeringen2017Apr,Nakamura2012May,Liang2015Aug,Moriya2014Aug}.
The 2D electron gas at surfaces and interfaces of cubic perovskite oxides attracts a large interest, mainly because of its high electron mobility and enhanced spin-orbit interaction that leads to strong spin-to-charge interconversion~\cite{Caviglia2010Mar,Lesne2016Dec,Trier2022Apr}. This makes oxide perovskites attractive materials for spintronics.  

This work is organised as follows. In Sec.~II. we introduce an effective Hamiltonian describing 2D electron gas with isotropic cubic Rashba SOC. In Sec.~III. we present semi-classical formulas describing linear and nonlinear Hall conductivity. Next, in Sec. IV.A. we present Berry curvature and BCD as a function of the in-plane magnetic field and parameters of the effective Hamiltonian. Sec. IV.B. contains detailed discussion on the behaviour of Hall conductivity in the in-plane magnetic field. Finally, Sec. V. contains the summary and final remarks.

\section{2D electron gas with {$\mathbf{k}$}-cubed Rashba SOC}

We consider effective low-energy Hamiltonian of 2D light-electrons (2D heavy-holes) with isotropic cubic form of Rashba SOC~\cite{Schliemann2005Feb,Liu2008Mar,vanHeeringen2017Apr}, that takes the following form:
\begin{equation}
\label{eq:hamiltonian}
	H= \frac{\hbar^2 k^2}{2m} {\hat{\sigma}}_0 + i \alpha(k_-^3 {\hat{\sigma}}_+ - k_+^3 {\hat{\sigma}}_-) + \mathbf{\Delta}
 \cdot \hat{\mathbf{s}} ,
\end{equation}
where $k^{2} = k_{x}^{2} + k_{y}^{2}$, and $m$ is the effective mass that in the systems under consideration is described by the following formula:
\begin{equation}
\label{eq:masa}
	m = m_0 \left( \gamma_1 + \gamma_2 - \frac{256\gamma_2^2}{3\pi^2(3\gamma_1+10\gamma_2)} \right)^{-1},
\end{equation}
with $m_{0}$ denoting the electron rest mass, and $\gamma_{1,2}$ standing for the Luttinger parameters.
The second term of Eq.~(\ref{eq:hamiltonian}) describes isotropic cubic Rashba SOC, where  $k_\pm \equiv (k_x \pm ik_y)$, $\hat{\sigma}_\pm \equiv \frac{1}{2}(\hat{\sigma}_x \pm i\hat{\sigma}_y)$ and $\hat{\sigma}_{0,x,y,z}$ are the identity and Pauli matrices. The Rashba parameter $\alpha$ reads:
\begin{equation}
\label{eq:alpha}
	\alpha = \frac{512eFL_z^4\gamma_2^2}{9\pi^6(3\gamma_1+10\gamma_2)(\gamma_1-2\gamma_2)},
\end{equation}
where $L_z$ and $F$ are the quantum well width and field strength, respectively. 
The last term in Hamiltonian~(\ref{eq:hamiltonian}) is a Zeeman-like term that describes the coupling of electron spin, $\mathbf{s}$, with magnetization and/or external magnetic field, that for keeping the generality is denoted here as $\mathbf{\Delta}$ (note that $\mathbf{\Delta}$ is defined in the energy units). The spin operators, $\hat{\mathbf{s}} = (\hat{s}_x, \hat{s}_y, \hat{s}_z)$, in this model are defined as follows~\cite{Liu2008Mar}:
\begin{subequations}
\begin{align}
	&\hat{s}_x = -s_0 k_y \hat{\sigma}_0 + s_1 (k_{-}^2 \hat{\sigma}_{+} + k_{+}^2 \hat{\sigma}_{-}),\\
	&\hat{s}_y = s_0 k_x \hat{\sigma}_0 + i s_{1} (k_{-}^2 \hat{\sigma}_{+} - k_{+}^2 \hat{\sigma}_{-}),\\
	&\hat{s}_z = \frac{3}{2}\hat{\sigma}_z,
\end{align}
\end{subequations}
and their forms are a consequence of two canonical transformations that need to be performed on Luttinger Hamiltonian to obtain the effective Hamiltonian~(\ref{eq:hamiltonian}). The coefficients $s_{0,1}$ are defined as:
\begin{equation}
\label{eq:s0}
	s_0 = \frac{512 eF L_z^4 \gamma_2 m_0}{9\pi^6\hbar^2 (3\gamma_1 + 10\gamma_2)(\gamma_1 - 2\gamma_2)}
\end{equation}
and
\begin{equation}
\label{eq:s1}
	s_1 = \left[ \frac{3}{4\pi^2} - \frac{256\gamma_2^2}{3\pi^4(3\gamma_1 + 10\gamma_2)^2} \right] L_z^2 .
\end{equation}

The material parameters that characterise different systems, have been collected for example in~\cite{Karwacki_PRB_2018}. Here, following~\cite{Liu2008Mar}  we chose the parameters $\gamma_1=7$, $\gamma_2=1.9$, $eF=5\cdot 10^{6}$~eV/m and $L_z=8.3$~nm.
As the necessary condition to obtain a nonzero Berry curvature is to have a nonzero z-component of $\mathbf{\Delta}$ (i.e., a nonzero macroscopic magnetization oriented out of plane of 2DEG or out of plane component of magnetic field), whereas to induce and tune the Berry curvature dipole we need in addition an in-plane  component of magnetic field, we introduce the following generalized notation: $\mathbf{\Delta} = (B_x, B_y, \Delta_{z})$. Here $B_{x,y}$ denote the external in-plane components of magnetic field in the energy units, and $\Delta_{z} = B_{z}$ or $\Delta_{z} = M_{z}$, where $B_{z}$  and $M_{z}$ denote the z-component of external magnetic field and the out-of plane macroscopic magnetization, respectively.

\begin{figure*}[t!]
    \centering
\includegraphics[width=0.93\textwidth]{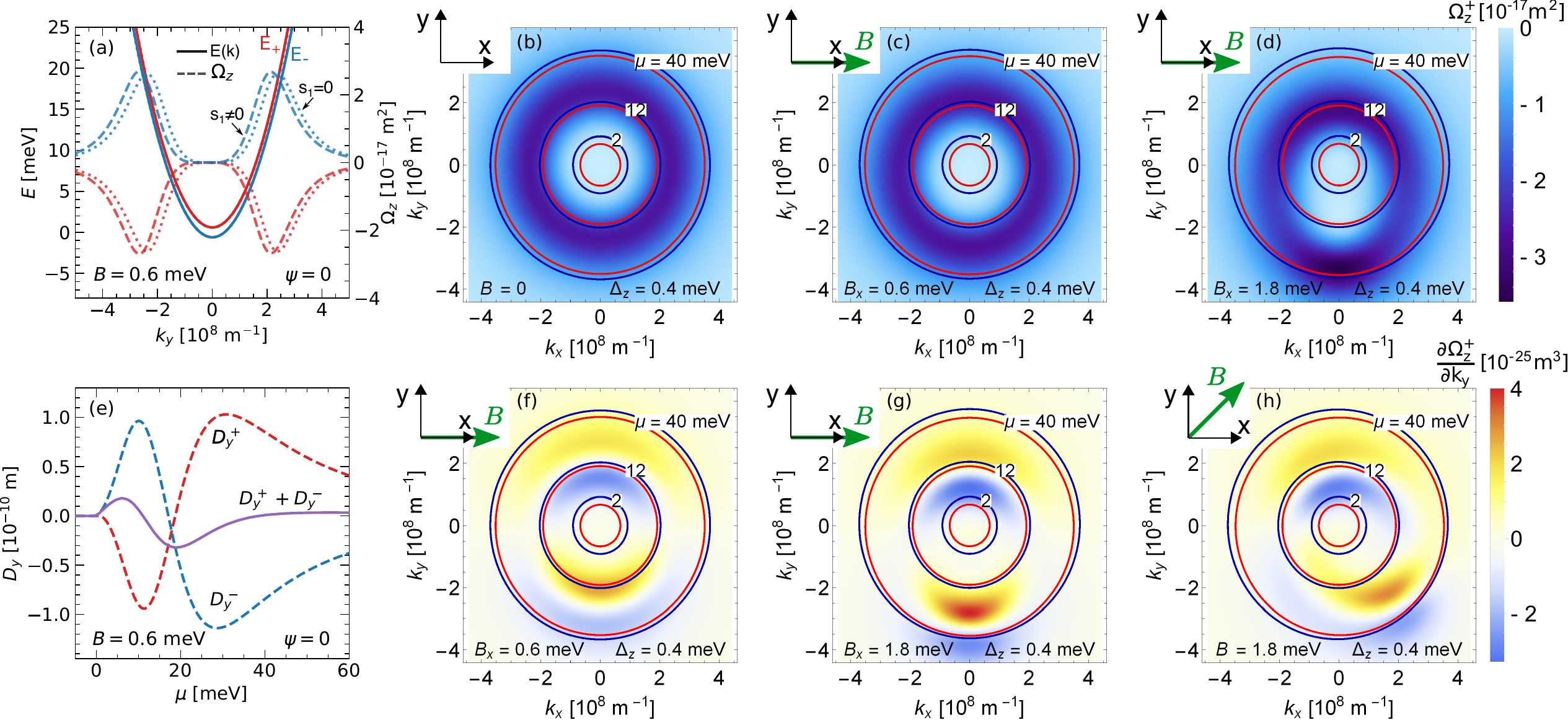}
\caption{Berry curvature (a-d) and Berry curvature dipole (e), as well as derivative of the Berry curvature with respect to $k_x$ plotted in $k$-space (f,h) for indicated values of $\Delta_{z}$ and $\mathbf{B}$. The other parameters are: $\gamma_1=7$, $\gamma_2=1.9$, $eF=5\cdot10^6$ eV/m, $L_z=8.3$~nm. }
    \label{fig:Fig1}
\end{figure*}
%

\section{Hall conductivity}
Within the semi-classical picture, the charge current density driven by ac longitudinal electric field can be described -- in the constant relaxation time approximation and up to the second order in the electric field -- by the following formula:
\begin{equation}
j_{\alpha} = {\mathrm{Re}}\{ j_{\alpha}^{0} + j_{\alpha}^{1} \exp(i\omega t) + j_{\alpha}^{2} \exp(i2\omega t)\},
\end{equation}
where the first term describes the rectification effect,
\begin{eqnarray}
j_{\alpha}^{0} = - \frac{e^{3} E_{\beta} E_{\gamma}^{\ast}}{2 \hbar^{2}}  \frac{\tau}{1 + i \omega \tau} \sum_{l} \int \frac{d^{2} \mathbf{k}}{(2\pi)^{2}} \left( \epsilon_{\alpha \gamma \delta} \Omega_{\delta}^{l} \partial_{\mathbf{k}}^{\beta}\right. \nonumber\\
\left.+ \tau v_{\alpha}^{l} \partial_{\mathbf{k}}^{\gamma}\partial_{\mathbf{k}}^{\beta} \right) f_{0}^{l},
\end{eqnarray}
the second term describes the linear response,
\begin{equation}
j_{\alpha}^{1} = -\frac{e^2}{\hbar} E_{\beta}\sum_{l} \int \frac{d^{2} \mathbf{k}}{(2\pi)^{2}} \left( \varepsilon_{\alpha \beta \gamma} \Omega_{\gamma}^{l} + \frac{\tau v_{\alpha}^{l}}{1 + i \omega \tau} \partial_{\mathbf{k}}^{\beta}\right) f_{0}^{l},
\end{equation}
and the last term is the second harmonic response,
\begin{eqnarray}
j_{\alpha}^{2} = - \frac{e^{3} E_{\beta} E_{\gamma}}{2 \hbar^{2}} \frac{\tau}{1 + i \omega \tau} \sum_{l} \int \frac{d^{2} \mathbf{k}}{(2\pi)^{2}} \left(\epsilon_{\alpha \gamma \delta} \Omega_{\delta}^{l} \partial_{\mathbf{k}}^{\beta}\right. \nonumber\\
\left.+ \frac{\tau v_{\alpha}^{l}}{1 + i 2\omega\tau} \partial_{\mathbf{k}}^{\gamma} \partial_{\mathbf{k}}^{\beta}  \right) f_{0}^{l}.
\end{eqnarray}
In the above expressions $l$ numerates the sub-bands, and $\{\alpha, \beta, \gamma, \delta\}= \{ x, y, z\}$ and $\Omega_{\alpha}^{l}$ is the $\alpha$-th component of the Berry Curvature calculated for the $l$-th band 
\begin{equation}
\Omega_{\gamma}^{l} = \nabla_{\mathbf{{k}}} \times \bm{\Lambda}_{l}({\mathbf{k}}) 
\end{equation}
where $\bm{\Lambda}_{l} = i \langle \psi_{l} | \nabla_{\mathbf{k}} | \psi_{l} \rangle$ is the Berry connection.

Accordingly, in the dc limit, one can write:
\begin{eqnarray}
j_{\alpha} = -\frac{e^2}{\hbar} E_{\beta}\sum_{l} \int \frac{d^{2} \mathbf{k}}{(2\pi)^{2}} \Big(\varepsilon_{\alpha \beta \gamma} \Omega_{\gamma}^{l} + \tau v_{\alpha}^{l} \partial_{\mathbf{k}}^{\beta}\Big.\hspace{2.3cm}\nonumber\\
+ \Big.\frac{e E_{\gamma} \tau}{\hbar} \left( \varepsilon_{\alpha \gamma \delta} \Omega_{\delta}^{l} \partial_{\mathbf{k}}^{\beta} + \tau v_{\alpha}^{l} \partial_{\mathbf{k}}^{\gamma} \partial_{\mathbf{k}}^{\beta}\right) \Big) f_{0}^{l}.\hspace{0.8cm}
\end{eqnarray}
Thus, the Hall conductivity can be written as follows: 
\begin{equation}
\label{eq:sig_alphabeta}
\sigma_{\alpha \beta} = \sigma_{\alpha \beta}^{I} + \sigma_{\alpha \beta}^{II} + \chi_{\alpha \beta \gamma}^{I} E_{\gamma} + \chi_{\alpha \beta \gamma}^{II} E_{\gamma}.
\end{equation}
Here, $\sigma_{\alpha \beta}^{I,II}$ are the components, that are well known from the theory of AHE. Thus, the first term describes the contributions from the electronic states at the Fermi level (dissipative term), that in the simplest form~\footnote{We do not consider here the  skew-scattering and side jump processes.} reads:
\begin{equation}
\sigma_{\alpha \beta}^{I} = -\frac{e^2}{\hbar} \tau \sum_{l} \int \frac{d^{2} \mathbf{k}}{(2\pi)^{2}} v_{\alpha}^{l} \partial_{\mathbf{k}}^{\beta} f_{0}^{l},
\end{equation}
whereas the second term describes the contribution from the electronic states below the Fermi level, i.e., the Fermi sea (non-dissipative) component. This term is fully intrinsic and determined by the Berry curvature:
\begin{equation}
\sigma_{\alpha \beta}^{II} = -\frac{e^2}{\hbar} \sum_{l} \int \frac{d^{2} \mathbf{k}}{(2\pi)^{2}} \varepsilon_{\alpha \beta \gamma} \Omega_{\gamma}^{l} f_{0}^{l}, 
\end{equation}

The third and fourth terms of Eq.~(\ref{eq:sig_alphabeta}) describe nonlinear Hall response. The third term is fully semi-classical, with electric susceptibility heaving the form:
\begin{equation}
\chi_{\alpha \beta \gamma}^{I} = - \ \frac{e^{2}}{\hbar} \frac{e \tau^{2}}{\hbar} \sum_{l} \int \frac{d^{2} \mathbf{k}}{(2\pi)^{2}} v_{\alpha}^{l} \partial_{\mathbf{k}}^{\gamma} \partial_{\mathbf{k}}^{\beta} f_{0}^{l},
\end{equation}
whereas the fourth term has quantum origin and the corresponding electric susceptibility takes the form:
\begin{align}
\chi_{\alpha \beta \gamma}^{II} &= - \ \frac{e^{2}}{\hbar} \frac{e \tau}{\hbar}\sum_{l} \int \frac{d^{2} \mathbf{k}}{(2\pi)^{2}} \varepsilon_{\alpha \gamma \delta} \Omega_{\delta}^{l} \partial_{\mathbf{k}}^{\beta} f_{0}^{l}\nonumber\\ &= -\frac{e^{2}}{\hbar} \frac{e \tau}{\hbar}\sum_{l} \varepsilon_{\alpha \gamma \delta} D_{\gamma \delta}^{l},
\end{align}
where $D_{\gamma \delta}$ is the Berry curvature dipole which in 2D systems (for which Berry curvature has only $\hat{z}$-component) reads:
\begin{equation}
\label{eq:D}
D_{\alpha \beta}^{l} \overset{2D}{=} D_{\alpha}^{l} = \int \frac{d^{2} \mathbf{k}}{(2\pi)^{2}} f_{0}^{l}\,\partial_{\mathbf{k}}^{\alpha} \Omega_{z}^{l}.
\end{equation}

\section{Results}


\subsection{Berry Curvature and Berry Curvature Dipole}

The eigenvalues, $E_{l=\pm}$,  of Hamiltonian~(\ref{eq:hamiltonian}) take the form:
\begin{equation}
    E_\pm = \frac{k^2 \hbar^2}{2 m}
    + s_0 \left(B_y k_x-B_x k_y\right)
    \pm \frac{\xi(\mathbf{k})}{2} ,
\end{equation}
and the corresponding eigenfunctions are:
\begin{equation}
\Psi_{l=\pm} = \begin{pmatrix}
    \pm i \Phi(\mathbf{k})\\
    1
    \end{pmatrix}.
\end{equation}
In the above expressions we introduced the parameters: 
{\small{$\xi(\mathbf{k}) = \left[9 \Delta_z^2 + 4 k^4 \left(B^2 s_1^2 + k^2\alpha^2 - 2 \alpha  (B_y k_x - B_x k_y)s_1\right) \right]^{1/2}$}} and  $\Phi(\mathbf{k}) = \frac{ \pm 3\Delta_z + \xi(\mathbf{k}) }{2 k_+^2 \left(\alpha k_+ + i (B_x - B_y)s_1\right)}$.
%
%
Accordingly, the Berry curvature corresponding to the  subbands $E_{\pm}$  has the following form:
\begin{equation}
\label{eq:isoCubRSOI_BC}
	\Omega^\pm_z = \mp \frac{6 \Delta_z k^2}{\left(\xi(\mathbf{k})\right)^{3}} \left[4 B^2 s_1^2+12 \alpha (B_x k_y-B_yk_x) s_1 + 9 \alpha^2 k^2\right] ,
\end{equation}
where $B^2=B_x^2 + B_y^2$ denotes the amplitude of the in-plane component of external magnetic field. In addition, one usually defines $(B_{x}/B, B_{y}/B) = (\cos\psi, \sin\psi)$, where $\psi$ determines the orientation of vector $\mathbf{B}$ with respect to the $\hat{\mathbf{x}}$-axis.

Both eigenvalues and Berry curvature are presented in Figs.~\ref{fig:Fig1}(a)-(d). Fig.~\ref{fig:Fig1}(a) shows the cross sections of energy bands $E_{\pm}$ (solid lines) and Berry curvatures corresponding to these eigenvalues. The Berry curvatures have also been plotted in the case, when the coefficient $s_{1}$ is taken into account in the definition of spin operators (dotted lines). Figures ~\ref{fig:Fig1}(b)-(d) present selected constant-energy contours and density plots for Berry curvature, $\Omega_{z}^{+}$, for a fixed value of $\Delta_{z}$ and for different amplitudes of the in-plane magnetic field, $B$. One can see that the nonzero in-plane magnetic field, $B$, leads to a Fermi contours anisotropy in the k-space by shifting the energy subbands (e.g., in the $k_x$ direction when the in-plane magnetic field is oriented along the $y$ direction). For higher energies, where the spin-orbit coupling also plays an important role, the $B$-field leads to a degeneracy of the subbands. Obviously, the Berry curvature reveals the highest signal in the crossing points. Moreover, the positions of positive and negative Berry curvature hot spots are also more pronounced at higher energies, and their position in the k-space is controlled by the in-plane orientation of the magnetic field (as indicated in~Fig.\ref{fig:Fig1}(f)-(h)). Figure~\ref{fig:Fig1}(e) presents the Berry curvature dipole calculated for $E_{+}$ and $E_{-}$ energy branches, respectively, as well as their superposition.  Importantly, depending on which component, $D_{y}^{+}$ or $D_{y}^{-}$, dominates the sum can change the sign by changing the chemical potential.

%

\subsection{Hall conductivity}

\begin{figure}[t!]
    \centering
\includegraphics[width=0.99\columnwidth]{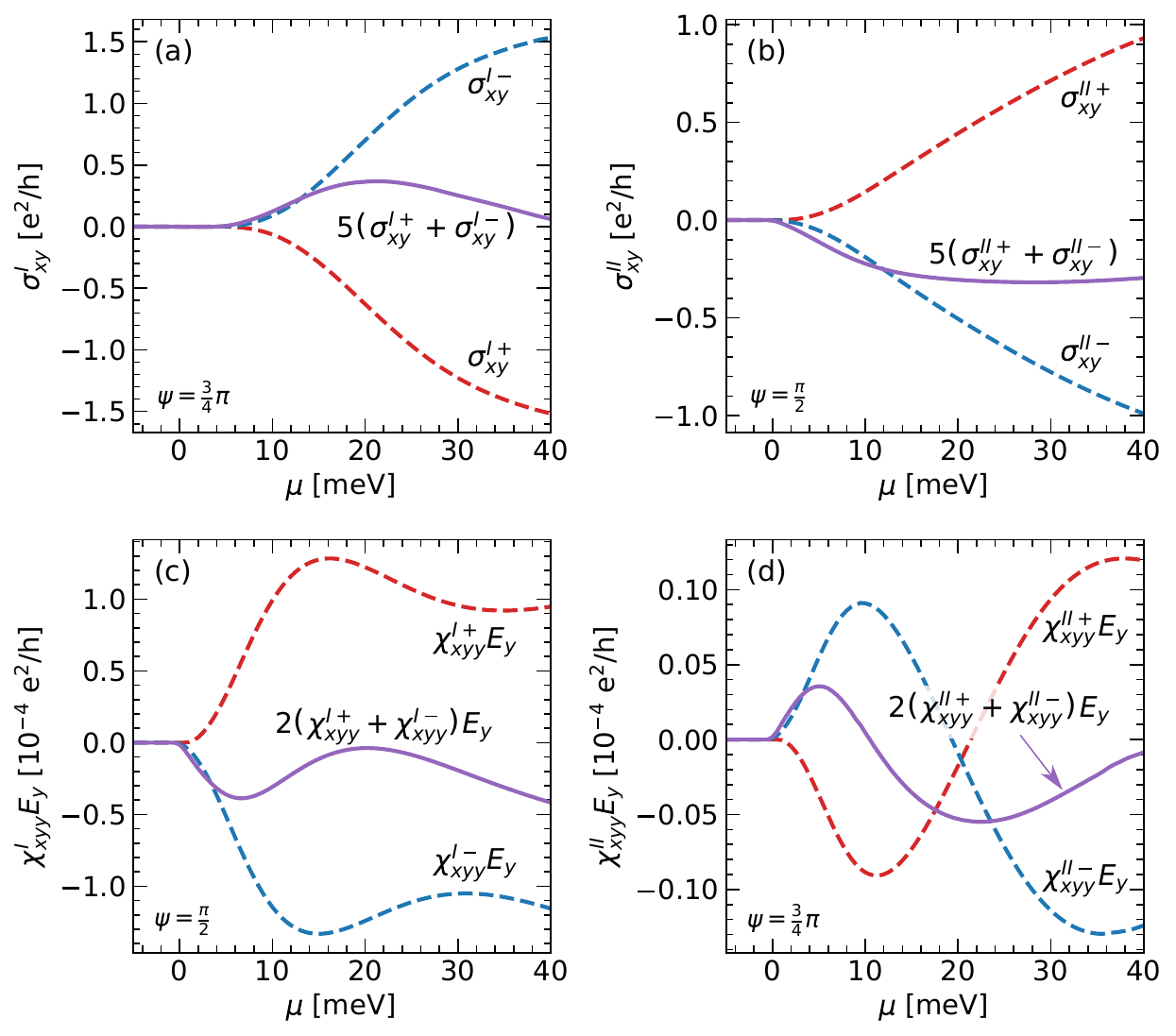}
    \caption{Four components of the Hall conductivity: $\sigma_{xy}^{I}$ (a), $\sigma_{xy}^{II}$ (b), $\chi_{xyy}^{I} E_{y}$ (c), and $\chi_{xyy}^{II} E_{y}$ (d) plotted as a function of chemical potential $\mu$ and for the indicated orientations of magnetic field  (i.e., for the specific values of the angle $\psi$). For all components, the contributions from each of the subbands are presented.  Here $\Delta_z=0.4$~meV, $B=1.8$~meV, $eE_y=1$~eV/m, $\Gamma=5\cdot10^{-5}$~eV, the other parameters are the same as  in Fig.~\ref{fig:Fig1}.}
    \label{fig:Fig2}
\end{figure}

\begin{figure*}[t!]
    \centering
\includegraphics[width=2.0\columnwidth]{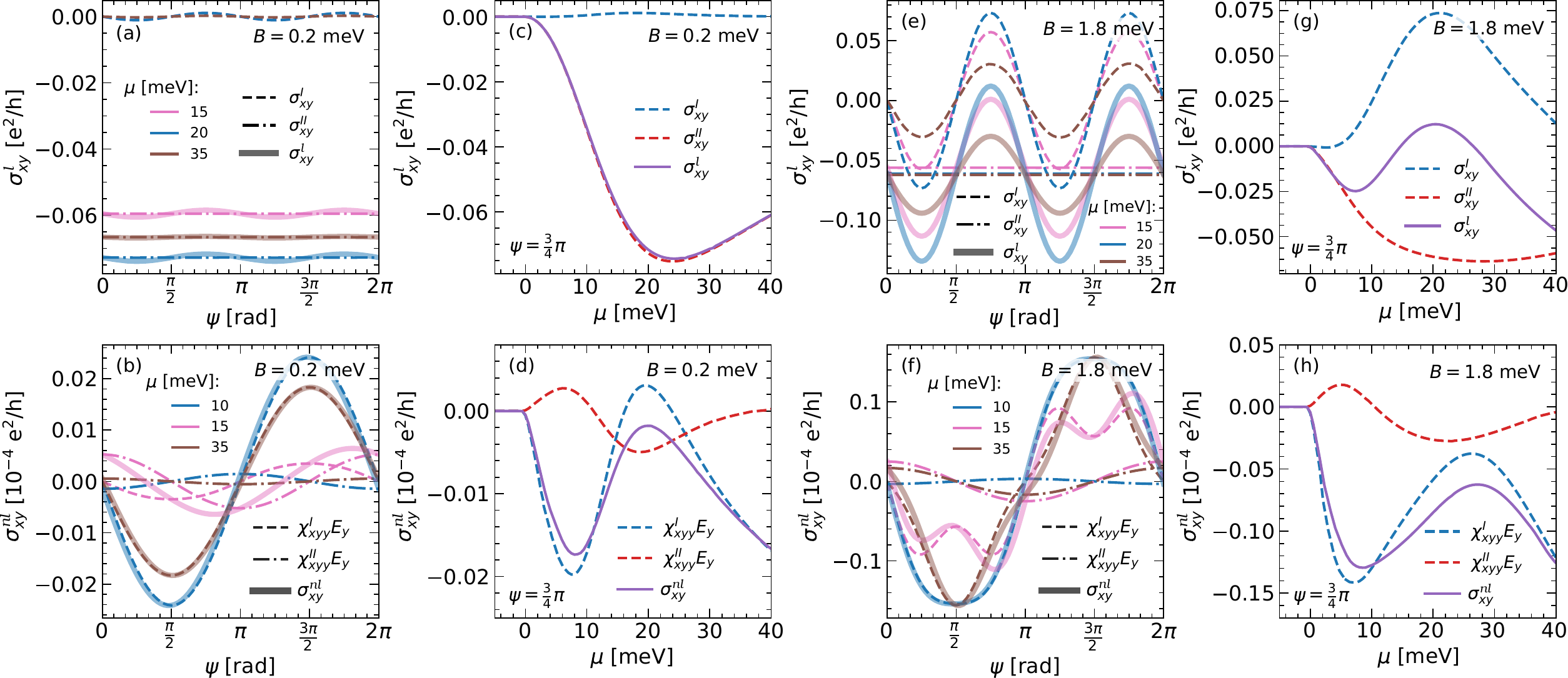}
    \caption{Linear and nonlinear Hall conductivity as a function of angle, $\psi$, defining the orientation of in-plane magnetic field (a),(b),(e)(f) and as a function of the chemical potential, $\mu$, (c),(d),(g),(h). The parameters are the same as in Fig.~\ref{fig:Fig1} and Fig.~\ref{fig:Fig2} unless otherwise indicated on the plots. }
    \label{fig:Fig3}
\end{figure*}

Now, without loss of generality, we assume that the external electric field is oriented in the $y$-direction. Accordingly, the Hall response is described by the off-diagonal conductivity $\sigma_{xy} = \sigma_{xy}^{I} + \sigma_{xy}^{II} + \chi_{xyy}^{I} E_{y} + \chi_{xyy}^{II} E_{y}$. The first two terms describe the linear, with respect to $E_{y}$, system response, i.e., $\sigma_{xy}^{l} = \sigma_{xy}^{I} + \sigma_{xy}^{II}$, whereas the last two terms describe the nonlinear (second-order) response, i.e., $\sigma_{xy}^{nl} = (\chi_{xyy}^{I} + \chi_{xyy}^{II}) E_{y}$. All these components are presented in Fig.~\ref{fig:Fig2} (where the contributions from each of the subbands are also plotted) and in Fig.~\ref{fig:Fig3}, where all conductivity components are presented as a function of the angle, $\psi$, and chemical potential, $\mu$, for two amplitudes of the in-plane magnetic field, i.e., for $B = 0.2$~meV and $B = 1.8$~meV. These two  values of B  represent two regimes: $B < \Delta_{z}$, and $B > \Delta_{z}$, respectively.

When $B < \Delta_{z}$, the intrinsic component, induced by the Berry curvature, dominates the linear system response. The component $\sigma_{xy}^{II}$ does not depend on the orientation of in-plane magnetic field (described by the angle $\psi$, defined as the angle between {$\hat{\textbf{x}}$}-direction and $\mathbf{B}$-vector), whereas the component $\sigma_{xy}^{I}$ is a periodic function of $\psi$, with the oscillation period equal to $\pi$. In turn, the nonlinear conductivity, $\sigma_{xy}^{nl}$ reveals $\psi$-dependent oscillations, with the oscillation period of $2\pi$. The oscillation period $2\pi$ is a signature of nonlinear with respect to the external electric field system response. Opposite to $\sigma_{xy}^{l}$, the topological term originating from the Berry curvature dipole is smaller than the nonlinear contribution originating from the states at the Fermi level. Both $\chi_{xyy}^{I}$ and $\chi_{xyy}^{II}$ are non-monotonous functions of chemical potential. In Fig.~\ref{fig:Fig3}(d), the nonlinear conductivity, $\sigma_{xy}^{nl}$, and its two components are plotted s a function of $\mu$ for $\psi = 3\pi/4$. From this plot, one can see that 
$\chi_{xyy}^{I}$ at first decreases with increasing value of $\mu$, reaches minimum around 9 meV and then increases, crosses 0, takes maximum around $\mu = 20$ meV, and next it decreases. In turn, $\chi_{xyy}^{II}$ is originally positive and takes maximum around $\mu = 9$ meV, then it decreases, crosses zero and takes minimum at $\mu = 20$meV    vanishes with further increase in $\mu$. As a result, $\sigma_{xy}^{nl}$ is a negative function having a local minimum around $\mu = 9$ meV and a local maximum around $\mu = 20$ meV.

When $B > \Delta_{z}$, the two components contributing to $\sigma_{xy}^{l}$, i.e., $\sigma_{xy}^{I}$ and $\sigma_{xy}^{II}$, are of the same order. Accordingly, $\sigma_{xy}^{l}$ reveals well defined oscillations with a $\pi$-periodicity. The component $\sigma_{xy}^{nl}$ has periodicity of $2\pi$. In addition, $\chi_{xyy}^{I}$ has a local maximum at $\psi = \pi/2$ and a local minimum at $\psi = 3\pi/2$, which indicates the presence of not only B linear but also B cubed terms. This means that when $\mathbf{B}$ is strong enough, not only the bilinear term, $ B_{x}E_{y}$, but also $B^{3}_{x}E_{y}$ term should be taken into account to describe properly $\sigma_{xy}^{nl}$. For $\psi = 3\pi/4$ the nonlinear contribution $\sigma_{xy}^{nl}$ is again a negative function, having a local minimum around $\mu = 9$ meV and a local maximum around $\mu = 20$ meV. 
We note, that
we have presented here numerical data only for the chemical potentials up to 40 meV, as the effective Hamiltonian given by Eq.~(\ref{eq:hamiltonian}) is valid only for small particle density (small wave numbers)~\cite{Schliemann2005Feb,Liu2008Mar}.

The symmetric part of the off-diagonal conductivity, $\Delta_{\scriptscriptstyle{S}} \sigma_{xy} = \left[ \sigma_{xy} \displaystyle{(E_y=E)} + \sigma_{xy} \displaystyle{(E_{y}=-E)}\right]/2 = \sigma_{xy}^{l}$ at lower in-plane magnetic field is determined by the topological component. The behaviour of $\sigma_{xy}^{II}$ for isotropic cubic Rashba model has been discussed in detail, for example, in Ref.~\cite{Krzyzewska2018}). In turn, when the amplitude of the in-plane magnetic field increases, the component $\sigma_{xy}^{I}$ becomes more visible and starts to dominate.

Importantly the antisymmetric part of the Hall conductivity $\Delta_{\scriptscriptstyle{A}}\sigma_{xy} = \left[\sigma_{xy}{\displaystyle{(E_y=E)}} - \sigma_{xy}{\displaystyle{(E_{y}=-E)}}\right]/2 = \sigma_{xy}^{nl}$ can be strongly modified not only by the orientation and amplitude of the in-plane magnetic field, but also by the change of chemical potential due to doping or gating. We have shown that the term related to the Berry curvature dipole is not sufficient to properly describe the behaviour of nonlinear system response, as the susceptibilities $\chi_{xyy}^{I}$ and $\chi_{xyy}^{II}$ can have opposite signs. Moreover, for larger in-plane magnetic fields, $\chi_{xyy}^{I}$  deviates from linear in $\mathbf{B}$ functional dependence.

\section{Discussion and Summary}

In this paper we have analysed theoretically the linear and nonlinear Hall effects in 2D systems within the isotropic k-cubed Rashba model. The analytical results are supported by numerical ones.  We showed that an out-of-plane external magnetic field or magnetization lead to a nonzero Berry curvature in the system. We have also shown, that the Berry curvature can be tuned by in-plane magnetic field, and this may lead to the Berry curvature dipole. 

We have also calculated  the linear and nonlinear Hall response, and have shown that  in the linear  response regime, the conductivity is dominated by the component due to Berry curvature. In turn, the second-order correction to the Hall conductivity is dominated by the states at the Fermi level.

\section*{Acknowledgement}
This work has been supported by the Norwegian Financial Mechanism 2014--2021 under the Polish Norwegian Research Project NCN GRIEG 2Dtronics no. 2019/34/H/ST3/00515.


\bibliography{BCDbib.bib}

\begin{thebibliography}{49}%
\makeatletter
\providecommand \@ifxundefined [1]{%
 \@ifx{#1\undefined}
}%
\providecommand \@ifnum [1]{%
 \ifnum #1\expandafter \@firstoftwo
 \else \expandafter \@secondoftwo
 \fi
}%
\providecommand \@ifx [1]{%
 \ifx #1\expandafter \@firstoftwo
 \else \expandafter \@secondoftwo
 \fi
}%
\providecommand \natexlab [1]{#1}%
\providecommand \enquote  [1]{``#1''}%
\providecommand \bibnamefont  [1]{#1}%
\providecommand \bibfnamefont [1]{#1}%
\providecommand \citenamefont [1]{#1}%
\providecommand \href@noop [0]{\@secondoftwo}%
\providecommand \href [0]{\begingroup \@sanitize@url \@href}%
\providecommand \@href[1]{\@@startlink{#1}\@@href}%
\providecommand \@@href[1]{\endgroup#1\@@endlink}%
\providecommand \@sanitize@url [0]{\catcode `\\12\catcode `\$12\catcode `\&12\catcode `\#12\catcode `\^12\catcode `\_12\catcode `\%12\relax}%
\providecommand \@@startlink[1]{}%
\providecommand \@@endlink[0]{}%
\providecommand \url  [0]{\begingroup\@sanitize@url \@url }%
\providecommand \@url [1]{\endgroup\@href {#1}{\urlprefix }}%
\providecommand \urlprefix  [0]{URL }%
\providecommand \Eprint [0]{\href }%
\providecommand \doibase [0]{https://doi.org/}%
\providecommand \selectlanguage [0]{\@gobble}%
\providecommand \bibinfo  [0]{\@secondoftwo}%
\providecommand \bibfield  [0]{\@secondoftwo}%
\providecommand \translation [1]{[#1]}%
\providecommand \BibitemOpen [0]{}%
\providecommand \bibitemStop [0]{}%
\providecommand \bibitemNoStop [0]{.\EOS\space}%
\providecommand \EOS [0]{\spacefactor3000\relax}%
\providecommand \BibitemShut  [1]{\csname bibitem#1\endcsname}%
\let\auto@bib@innerbib\@empty
\bibitem [{\citenamefont {Bychkov}\ and\ \citenamefont {Rashba}(1984)}]{Bychkov1984}%
  \BibitemOpen
  \bibfield  {author} {\bibinfo {author} {\bibfnamefont {Y.~A.}\ \bibnamefont {Bychkov}}\ and\ \bibinfo {author} {\bibfnamefont {E.~I.}\ \bibnamefont {Rashba}},\ }\bibfield  {title} {\bibinfo {title} {{Properties of a 2D Electron Gas with Lifted Spectral Degeneracy}},\ }\href@noop {} {\bibfield  {journal} {\bibinfo  {journal} {JETP Letters}\ }\textbf {\bibinfo {volume} {39}},\ \bibinfo {pages} {78} (\bibinfo {year} {1984})}\BibitemShut {NoStop}%
\bibitem [{\citenamefont {Winkler}(2003)}]{Winkler2003}%
  \BibitemOpen
  \bibfield  {author} {\bibinfo {author} {\bibfnamefont {R.}~\bibnamefont {Winkler}},\ }\href {https://doi.org/10.1007/b13586} {\emph {\bibinfo {title} {{Spin{\ifmmode---\else\textemdash\fi}Orbit Coupling Effects in Two-Dimensional Electron and Hole Systems}}}}\ (\bibinfo  {publisher} {Springer},\ \bibinfo {address} {Berlin, Germany},\ \bibinfo {year} {2003})\BibitemShut {NoStop}%
\bibitem [{\citenamefont {Bihlmayer}\ \emph {et~al.}(2015)\citenamefont {Bihlmayer}, \citenamefont {Rader},\ and\ \citenamefont {Winkler}}]{Bihlmayer2015May}%
  \BibitemOpen
  \bibfield  {author} {\bibinfo {author} {\bibfnamefont {G.}~\bibnamefont {Bihlmayer}}, \bibinfo {author} {\bibfnamefont {O.}~\bibnamefont {Rader}},\ and\ \bibinfo {author} {\bibfnamefont {R.}~\bibnamefont {Winkler}},\ }\bibfield  {title} {\bibinfo {title} {{Focus on the Rashba effect}},\ }\href {https://doi.org/10.1088/1367-2630/17/5/050202} {\bibfield  {journal} {\bibinfo  {journal} {New J. Phys.}\ }\textbf {\bibinfo {volume} {17}},\ \bibinfo {pages} {050202} (\bibinfo {year} {2015})}\BibitemShut {NoStop}%
\bibitem [{\citenamefont {Manchon}\ \emph {et~al.}(2015)\citenamefont {Manchon}, \citenamefont {Koo}, \citenamefont {Nitta}, \citenamefont {Frolov},\ and\ \citenamefont {Duine}}]{Manchon_NatMat2015}%
  \BibitemOpen
  \bibfield  {author} {\bibinfo {author} {\bibfnamefont {A.}~\bibnamefont {Manchon}}, \bibinfo {author} {\bibfnamefont {H.~C.}\ \bibnamefont {Koo}}, \bibinfo {author} {\bibfnamefont {J.}~\bibnamefont {Nitta}}, \bibinfo {author} {\bibfnamefont {S.~M.}\ \bibnamefont {Frolov}},\ and\ \bibinfo {author} {\bibfnamefont {R.~A.}\ \bibnamefont {Duine}},\ }\bibfield  {title} {\bibinfo {title} {{New perspectives for Rashba spin--orbit coupling}},\ }\href {https://doi.org/10.1038/nmat4360} {\bibfield  {journal} {\bibinfo  {journal} {Nature Materials}\ }\textbf {\bibinfo {volume} {14}},\ \bibinfo {pages} {871} (\bibinfo {year} {2015})}\BibitemShut {NoStop}%
\bibitem [{\citenamefont {Bihlmayer}\ \emph {et~al.}(2022)\citenamefont {Bihlmayer}, \citenamefont {No{\"e}l}, \citenamefont {Vyalikh}, \citenamefont {Chulkov},\ and\ \citenamefont {Manchon}}]{Bihlmayer_NatRevPhys2022}%
  \BibitemOpen
  \bibfield  {author} {\bibinfo {author} {\bibfnamefont {G.}~\bibnamefont {Bihlmayer}}, \bibinfo {author} {\bibfnamefont {P.}~\bibnamefont {No{\"e}l}}, \bibinfo {author} {\bibfnamefont {D.~V.}\ \bibnamefont {Vyalikh}}, \bibinfo {author} {\bibfnamefont {E.~V.}\ \bibnamefont {Chulkov}},\ and\ \bibinfo {author} {\bibfnamefont {A.}~\bibnamefont {Manchon}},\ }\bibfield  {title} {\bibinfo {title} {{Rashba-like physics in condensed matter}},\ }\href {https://doi.org/10.1038/s42254-022-00490-y} {\bibfield  {journal} {\bibinfo  {journal} {Nature Reviews Physics}\ }\textbf {\bibinfo {volume} {4}},\ \bibinfo {pages} {642} (\bibinfo {year} {2022})}\BibitemShut {NoStop}%
\bibitem [{\citenamefont {Soumyanarayanan}\ \emph {et~al.}(2016)\citenamefont {Soumyanarayanan}, \citenamefont {Reyren}, \citenamefont {Fert},\ and\ \citenamefont {Panagopoulos}}]{Soumyanarayanan2016Nov}%
  \BibitemOpen
  \bibfield  {author} {\bibinfo {author} {\bibfnamefont {A.}~\bibnamefont {Soumyanarayanan}}, \bibinfo {author} {\bibfnamefont {N.}~\bibnamefont {Reyren}}, \bibinfo {author} {\bibfnamefont {A.}~\bibnamefont {Fert}},\ and\ \bibinfo {author} {\bibfnamefont {C.}~\bibnamefont {Panagopoulos}},\ }\bibfield  {title} {\bibinfo {title} {{Emergent phenomena induced by spin{\textendash}orbit coupling at surfaces and interfaces}},\ }\href {https://doi.org/10.1038/nature19820} {\bibfield  {journal} {\bibinfo  {journal} {Nature}\ }\textbf {\bibinfo {volume} {539}},\ \bibinfo {pages} {509} (\bibinfo {year} {2016})}\BibitemShut {NoStop}%
\bibitem [{\citenamefont {Feng}\ \emph {et~al.}(2017)\citenamefont {Feng}, \citenamefont {Shen}, \citenamefont {Yang}, \citenamefont {Wang}, \citenamefont {Zeng}, \citenamefont {Wu}, \citenamefont {Chintalapati},\ and\ \citenamefont {Chang}}]{Feng2017Sep}%
  \BibitemOpen
  \bibfield  {author} {\bibinfo {author} {\bibfnamefont {Y.~P.}\ \bibnamefont {Feng}}, \bibinfo {author} {\bibfnamefont {L.}~\bibnamefont {Shen}}, \bibinfo {author} {\bibfnamefont {M.}~\bibnamefont {Yang}}, \bibinfo {author} {\bibfnamefont {A.}~\bibnamefont {Wang}}, \bibinfo {author} {\bibfnamefont {M.}~\bibnamefont {Zeng}}, \bibinfo {author} {\bibfnamefont {Q.}~\bibnamefont {Wu}}, \bibinfo {author} {\bibfnamefont {S.}~\bibnamefont {Chintalapati}},\ and\ \bibinfo {author} {\bibfnamefont {C.-R.}\ \bibnamefont {Chang}},\ }\bibfield  {title} {\bibinfo {title} {{Prospects of spintronics based on 2D materials}},\ }\href {https://doi.org/10.1002/wcms.1313} {\bibfield  {journal} {\bibinfo  {journal} {WIREs Comput. Mol. Sci.}\ }\textbf {\bibinfo {volume} {7}},\ \bibinfo {pages} {e1313} (\bibinfo {year} {2017})}\BibitemShut {NoStop}%
\bibitem [{\citenamefont {Hirohata}\ \emph {et~al.}(2020)\citenamefont {Hirohata}, \citenamefont {Yamada}, \citenamefont {Nakatani}, \citenamefont {Prejbeanu}, \citenamefont {Di{\ifmmode\acute{e}\else\'{e}\fi}ny}, \citenamefont {Pirro},\ and\ \citenamefont {Hillebrands}}]{Hirohata2020Sep}%
  \BibitemOpen
  \bibfield  {author} {\bibinfo {author} {\bibfnamefont {A.}~\bibnamefont {Hirohata}}, \bibinfo {author} {\bibfnamefont {K.}~\bibnamefont {Yamada}}, \bibinfo {author} {\bibfnamefont {Y.}~\bibnamefont {Nakatani}}, \bibinfo {author} {\bibfnamefont {I.-L.}\ \bibnamefont {Prejbeanu}}, \bibinfo {author} {\bibfnamefont {B.}~\bibnamefont {Di{\ifmmode\acute{e}\else\'{e}\fi}ny}}, \bibinfo {author} {\bibfnamefont {P.}~\bibnamefont {Pirro}},\ and\ \bibinfo {author} {\bibfnamefont {B.}~\bibnamefont {Hillebrands}},\ }\bibfield  {title} {\bibinfo {title} {{Review on spintronics: Principles and device applications}},\ }\href {https://doi.org/10.1016/j.jmmm.2020.166711} {\bibfield  {journal} {\bibinfo  {journal} {J. Magn. Magn. Mater.}\ }\textbf {\bibinfo {volume} {509}},\ \bibinfo {pages} {166711} (\bibinfo {year} {2020})}\BibitemShut {NoStop}%
\bibitem [{\citenamefont {Hsieh}\ \emph {et~al.}(2009)\citenamefont {Hsieh}, \citenamefont {Xia}, \citenamefont {Qian}, \citenamefont {Wray}, \citenamefont {Dil}, \citenamefont {Meier}, \citenamefont {Osterwalder}, \citenamefont {Patthey}, \citenamefont {Checkelsky}, \citenamefont {Ong}, \citenamefont {Fedorov}, \citenamefont {Lin}, \citenamefont {Bansil}, \citenamefont {Grauer}, \citenamefont {Hor}, \citenamefont {Cava},\ and\ \citenamefont {Hasan}}]{Hsieh2009Aug}%
  \BibitemOpen
  \bibfield  {author} {\bibinfo {author} {\bibfnamefont {D.}~\bibnamefont {Hsieh}}, \bibinfo {author} {\bibfnamefont {Y.}~\bibnamefont {Xia}}, \bibinfo {author} {\bibfnamefont {D.}~\bibnamefont {Qian}}, \bibinfo {author} {\bibfnamefont {L.}~\bibnamefont {Wray}}, \bibinfo {author} {\bibfnamefont {J.~H.}\ \bibnamefont {Dil}}, \bibinfo {author} {\bibfnamefont {F.}~\bibnamefont {Meier}}, \bibinfo {author} {\bibfnamefont {J.}~\bibnamefont {Osterwalder}}, \bibinfo {author} {\bibfnamefont {L.}~\bibnamefont {Patthey}}, \bibinfo {author} {\bibfnamefont {J.~G.}\ \bibnamefont {Checkelsky}}, \bibinfo {author} {\bibfnamefont {N.~P.}\ \bibnamefont {Ong}}, \bibinfo {author} {\bibfnamefont {A.~V.}\ \bibnamefont {Fedorov}}, \bibinfo {author} {\bibfnamefont {H.}~\bibnamefont {Lin}}, \bibinfo {author} {\bibfnamefont {A.}~\bibnamefont {Bansil}}, \bibinfo {author} {\bibfnamefont {D.}~\bibnamefont {Grauer}}, \bibinfo {author} {\bibfnamefont {Y.~S.}\ \bibnamefont {Hor}}, \bibinfo {author} {\bibfnamefont {R.~J.}\ \bibnamefont
  {Cava}},\ and\ \bibinfo {author} {\bibfnamefont {M.~Z.}\ \bibnamefont {Hasan}},\ }\bibfield  {title} {\bibinfo {title} {{A tunable topological insulator in the spin helical Dirac transport regime}},\ }\href {https://doi.org/10.1038/nature08234} {\bibfield  {journal} {\bibinfo  {journal} {Nature}\ }\textbf {\bibinfo {volume} {460}},\ \bibinfo {pages} {1101} (\bibinfo {year} {2009})}\BibitemShut {NoStop}%
\bibitem [{\citenamefont {Bernevig}\ and\ \citenamefont {Vafek}(2005)}]{Bernevig2005Jul}%
  \BibitemOpen
  \bibfield  {author} {\bibinfo {author} {\bibfnamefont {B.~A.}\ \bibnamefont {Bernevig}}\ and\ \bibinfo {author} {\bibfnamefont {O.}~\bibnamefont {Vafek}},\ }\bibfield  {title} {\bibinfo {title} {{Piezo-magnetoelectric effects in $p$-doped semiconductors}},\ }\href {https://doi.org/10.1103/PhysRevB.72.033203} {\bibfield  {journal} {\bibinfo  {journal} {Phys. Rev. B}\ }\textbf {\bibinfo {volume} {72}},\ \bibinfo {pages} {033203} (\bibinfo {year} {2005})}\BibitemShut {NoStop}%
\bibitem [{\citenamefont {Chernyshov}\ \emph {et~al.}(2009)\citenamefont {Chernyshov}, \citenamefont {Overby}, \citenamefont {Liu}, \citenamefont {Furdyna}, \citenamefont {Lyanda-Geller},\ and\ \citenamefont {Rokhinson}}]{Chernyshov2009Sep}%
  \BibitemOpen
  \bibfield  {author} {\bibinfo {author} {\bibfnamefont {A.}~\bibnamefont {Chernyshov}}, \bibinfo {author} {\bibfnamefont {M.}~\bibnamefont {Overby}}, \bibinfo {author} {\bibfnamefont {X.}~\bibnamefont {Liu}}, \bibinfo {author} {\bibfnamefont {J.~K.}\ \bibnamefont {Furdyna}}, \bibinfo {author} {\bibfnamefont {Y.}~\bibnamefont {Lyanda-Geller}},\ and\ \bibinfo {author} {\bibfnamefont {L.~P.}\ \bibnamefont {Rokhinson}},\ }\bibfield  {title} {\bibinfo {title} {{Evidence for reversible control of magnetization in a ferromagnetic material by means of spin{\textendash}orbit magnetic field}},\ }\href {https://doi.org/10.1038/nphys1362} {\bibfield  {journal} {\bibinfo  {journal} {Nat. Phys.}\ }\textbf {\bibinfo {volume} {5}},\ \bibinfo {pages} {656} (\bibinfo {year} {2009})}\BibitemShut {NoStop}%
\bibitem [{\citenamefont {Kato}\ \emph {et~al.}(2004{\natexlab{a}})\citenamefont {Kato}, \citenamefont {Myers}, \citenamefont {Gossard},\ and\ \citenamefont {Awschalom}}]{Kato2004Oct}%
  \BibitemOpen
  \bibfield  {author} {\bibinfo {author} {\bibfnamefont {Y.~K.}\ \bibnamefont {Kato}}, \bibinfo {author} {\bibfnamefont {R.~C.}\ \bibnamefont {Myers}}, \bibinfo {author} {\bibfnamefont {A.~C.}\ \bibnamefont {Gossard}},\ and\ \bibinfo {author} {\bibfnamefont {D.~D.}\ \bibnamefont {Awschalom}},\ }\bibfield  {title} {\bibinfo {title} {{Current-Induced Spin Polarization in Strained Semiconductors}},\ }\href {https://doi.org/10.1103/PhysRevLett.93.176601} {\bibfield  {journal} {\bibinfo  {journal} {Phys. Rev. Lett.}\ }\textbf {\bibinfo {volume} {93}},\ \bibinfo {pages} {176601} (\bibinfo {year} {2004}{\natexlab{a}})}\BibitemShut {NoStop}%
\bibitem [{\citenamefont {Sih}\ \emph {et~al.}(2005)\citenamefont {Sih}, \citenamefont {Myers}, \citenamefont {Kato}, \citenamefont {Lau}, \citenamefont {Gossard},\ and\ \citenamefont {Awschalom}}]{Sih2005Oct}%
  \BibitemOpen
  \bibfield  {author} {\bibinfo {author} {\bibfnamefont {V.}~\bibnamefont {Sih}}, \bibinfo {author} {\bibfnamefont {R.~C.}\ \bibnamefont {Myers}}, \bibinfo {author} {\bibfnamefont {Y.~K.}\ \bibnamefont {Kato}}, \bibinfo {author} {\bibfnamefont {W.~H.}\ \bibnamefont {Lau}}, \bibinfo {author} {\bibfnamefont {A.~C.}\ \bibnamefont {Gossard}},\ and\ \bibinfo {author} {\bibfnamefont {D.~D.}\ \bibnamefont {Awschalom}},\ }\bibfield  {title} {\bibinfo {title} {{Spatial imaging of the spin Hall effect and current-induced polarization in two-dimensional electron gases}},\ }\href {https://doi.org/10.1038/nphys009} {\bibfield  {journal} {\bibinfo  {journal} {Nat. Phys.}\ }\textbf {\bibinfo {volume} {1}},\ \bibinfo {pages} {31} (\bibinfo {year} {2005})}\BibitemShut {NoStop}%
\bibitem [{\citenamefont {Li}\ \emph {et~al.}(2014)\citenamefont {Li}, \citenamefont {van~'t Erve}, \citenamefont {Robinson}, \citenamefont {Liu}, \citenamefont {Li},\ and\ \citenamefont {Jonker}}]{Li2014Mar}%
  \BibitemOpen
  \bibfield  {author} {\bibinfo {author} {\bibfnamefont {C.~H.}\ \bibnamefont {Li}}, \bibinfo {author} {\bibfnamefont {O.~M.~J.}\ \bibnamefont {van~'t Erve}}, \bibinfo {author} {\bibfnamefont {J.~T.}\ \bibnamefont {Robinson}}, \bibinfo {author} {\bibfnamefont {Y.}~\bibnamefont {Liu}}, \bibinfo {author} {\bibfnamefont {L.}~\bibnamefont {Li}},\ and\ \bibinfo {author} {\bibfnamefont {B.~T.}\ \bibnamefont {Jonker}},\ }\bibfield  {title} {\bibinfo {title} {{Electrical detection of charge-current-induced spin polarization due to spin-momentum locking in Bi2Se3}},\ }\href {https://doi.org/10.1038/nnano.2014.16} {\bibfield  {journal} {\bibinfo  {journal} {Nat. Nanotechnol.}\ }\textbf {\bibinfo {volume} {9}},\ \bibinfo {pages} {218} (\bibinfo {year} {2014})},\ \Eprint {https://arxiv.org/abs/24561354} {24561354} \BibitemShut {NoStop}%
\bibitem [{\citenamefont {Sinova}\ \emph {et~al.}(2015)\citenamefont {Sinova}, \citenamefont {Valenzuela}, \citenamefont {Wunderlich}, \citenamefont {Back},\ and\ \citenamefont {Jungwirth}}]{Sinova_RevModPhys_SHE_2015}%
  \BibitemOpen
  \bibfield  {author} {\bibinfo {author} {\bibfnamefont {J.}~\bibnamefont {Sinova}}, \bibinfo {author} {\bibfnamefont {S.~O.}\ \bibnamefont {Valenzuela}}, \bibinfo {author} {\bibfnamefont {J.}~\bibnamefont {Wunderlich}}, \bibinfo {author} {\bibfnamefont {C.~H.}\ \bibnamefont {Back}},\ and\ \bibinfo {author} {\bibfnamefont {T.}~\bibnamefont {Jungwirth}},\ }\bibfield  {title} {\bibinfo {title} {Spin hall effects},\ }\href {https://doi.org/10.1103/RevModPhys.87.1213} {\bibfield  {journal} {\bibinfo  {journal} {Rev. Mod. Phys.}\ }\textbf {\bibinfo {volume} {87}},\ \bibinfo {pages} {1213} (\bibinfo {year} {2015})}\BibitemShut {NoStop}%
\bibitem [{\citenamefont {Kato}\ \emph {et~al.}(2004{\natexlab{b}})\citenamefont {Kato}, \citenamefont {Myers}, \citenamefont {Gossard},\ and\ \citenamefont {Awschalom}}]{Kato2004Dec}%
  \BibitemOpen
  \bibfield  {author} {\bibinfo {author} {\bibfnamefont {Y.~K.}\ \bibnamefont {Kato}}, \bibinfo {author} {\bibfnamefont {R.~C.}\ \bibnamefont {Myers}}, \bibinfo {author} {\bibfnamefont {A.~C.}\ \bibnamefont {Gossard}},\ and\ \bibinfo {author} {\bibfnamefont {D.~D.}\ \bibnamefont {Awschalom}},\ }\bibfield  {title} {\bibinfo {title} {{Observation of the Spin Hall Effect in Semiconductors}},\ }\href {https://doi.org/10.1126/science.1105514} {\bibfield  {journal} {\bibinfo  {journal} {Science}\ }\textbf {\bibinfo {volume} {306}},\ \bibinfo {pages} {1910} (\bibinfo {year} {2004}{\natexlab{b}})}\BibitemShut {NoStop}%
\bibitem [{\citenamefont {Wunderlich}\ \emph {et~al.}(2005)\citenamefont {Wunderlich}, \citenamefont {Kaestner}, \citenamefont {Sinova},\ and\ \citenamefont {Jungwirth}}]{Wunderlich2005Feb}%
  \BibitemOpen
  \bibfield  {author} {\bibinfo {author} {\bibfnamefont {J.}~\bibnamefont {Wunderlich}}, \bibinfo {author} {\bibfnamefont {B.}~\bibnamefont {Kaestner}}, \bibinfo {author} {\bibfnamefont {J.}~\bibnamefont {Sinova}},\ and\ \bibinfo {author} {\bibfnamefont {T.}~\bibnamefont {Jungwirth}},\ }\bibfield  {title} {\bibinfo {title} {{Experimental Observation of the Spin-Hall Effect in a Two-Dimensional Spin-Orbit Coupled Semiconductor System}},\ }\href {https://doi.org/10.1103/PhysRevLett.94.047204} {\bibfield  {journal} {\bibinfo  {journal} {Phys. Rev. Lett.}\ }\textbf {\bibinfo {volume} {94}},\ \bibinfo {pages} {047204} (\bibinfo {year} {2005})}\BibitemShut {NoStop}%
\bibitem [{\citenamefont {Valenzuela}\ and\ \citenamefont {Tinkham}(2006)}]{Valenzuela2006Jul}%
  \BibitemOpen
  \bibfield  {author} {\bibinfo {author} {\bibfnamefont {S.~O.}\ \bibnamefont {Valenzuela}}\ and\ \bibinfo {author} {\bibfnamefont {M.}~\bibnamefont {Tinkham}},\ }\bibfield  {title} {\bibinfo {title} {{Direct electronic measurement of the spin Hall effect}},\ }\href {https://doi.org/10.1038/nature04937} {\bibfield  {journal} {\bibinfo  {journal} {Nature}\ }\textbf {\bibinfo {volume} {442}},\ \bibinfo {pages} {176} (\bibinfo {year} {2006})}\BibitemShut {NoStop}%
\bibitem [{\citenamefont {Saitoh}\ \emph {et~al.}(2006)\citenamefont {Saitoh}, \citenamefont {Ueda}, \citenamefont {Miyajima},\ and\ \citenamefont {Tatara}}]{Saitoh2006May}%
  \BibitemOpen
  \bibfield  {author} {\bibinfo {author} {\bibfnamefont {E.}~\bibnamefont {Saitoh}}, \bibinfo {author} {\bibfnamefont {M.}~\bibnamefont {Ueda}}, \bibinfo {author} {\bibfnamefont {H.}~\bibnamefont {Miyajima}},\ and\ \bibinfo {author} {\bibfnamefont {G.}~\bibnamefont {Tatara}},\ }\bibfield  {title} {\bibinfo {title} {{Conversion of spin current into charge current at room temperature: Inverse spin-Hall effect}},\ }\href {https://doi.org/10.1063/1.2199473} {\bibfield  {journal} {\bibinfo  {journal} {Appl. Phys. Lett.}\ }\textbf {\bibinfo {volume} {88}},\ \bibinfo {pages} {182509} (\bibinfo {year} {2006})}\BibitemShut {NoStop}%
\bibitem [{\citenamefont {Wunderlich}\ \emph {et~al.}(2010)\citenamefont {Wunderlich}, \citenamefont {Park}, \citenamefont {Irvine}, \citenamefont {Z{\ifmmode\hat{a}\else\^{a}\fi}rbo}, \citenamefont {Rozkotov{\ifmmode\acute{a}\else\'{a}\fi}}, \citenamefont {Nemec}, \citenamefont {Nov{\ifmmode\acute{a}\else\'{a}\fi}k}, \citenamefont {Sinova},\ and\ \citenamefont {Jungwirth}}]{Wunderlich2010Dec}%
  \BibitemOpen
  \bibfield  {author} {\bibinfo {author} {\bibfnamefont {J.}~\bibnamefont {Wunderlich}}, \bibinfo {author} {\bibfnamefont {B.-G.}\ \bibnamefont {Park}}, \bibinfo {author} {\bibfnamefont {A.~C.}\ \bibnamefont {Irvine}}, \bibinfo {author} {\bibfnamefont {L.~P.}\ \bibnamefont {Z{\ifmmode\hat{a}\else\^{a}\fi}rbo}}, \bibinfo {author} {\bibfnamefont {E.}~\bibnamefont {Rozkotov{\ifmmode\acute{a}\else\'{a}\fi}}}, \bibinfo {author} {\bibfnamefont {P.}~\bibnamefont {Nemec}}, \bibinfo {author} {\bibfnamefont {V.}~\bibnamefont {Nov{\ifmmode\acute{a}\else\'{a}\fi}k}}, \bibinfo {author} {\bibfnamefont {J.}~\bibnamefont {Sinova}},\ and\ \bibinfo {author} {\bibfnamefont {T.}~\bibnamefont {Jungwirth}},\ }\bibfield  {title} {\bibinfo {title} {{Spin Hall Effect Transistor}},\ }\href {https://doi.org/10.1126/science.1195816} {\bibfield  {journal} {\bibinfo  {journal} {Science}\ }\textbf {\bibinfo {volume} {330}},\ \bibinfo {pages} {1801} (\bibinfo {year} {2010})}\BibitemShut {NoStop}%
\bibitem [{\citenamefont {He}\ \emph {et~al.}(2018)\citenamefont {He}, \citenamefont {Zhang}, \citenamefont {Zhu}, \citenamefont {Liu}, \citenamefont {Wang}, \citenamefont {Yu}, \citenamefont {Vignale},\ and\ \citenamefont {Yang}}]{He2018May}%
  \BibitemOpen
  \bibfield  {author} {\bibinfo {author} {\bibfnamefont {P.}~\bibnamefont {He}}, \bibinfo {author} {\bibfnamefont {S.~S.-L.}\ \bibnamefont {Zhang}}, \bibinfo {author} {\bibfnamefont {D.}~\bibnamefont {Zhu}}, \bibinfo {author} {\bibfnamefont {Y.}~\bibnamefont {Liu}}, \bibinfo {author} {\bibfnamefont {Y.}~\bibnamefont {Wang}}, \bibinfo {author} {\bibfnamefont {J.}~\bibnamefont {Yu}}, \bibinfo {author} {\bibfnamefont {G.}~\bibnamefont {Vignale}},\ and\ \bibinfo {author} {\bibfnamefont {H.}~\bibnamefont {Yang}},\ }\bibfield  {title} {\bibinfo {title} {{Bilinear magnetoelectric resistance as a probe of three-dimensional spin texture in topological surface states}},\ }\href {https://doi.org/10.1038/s41567-017-0039-y} {\bibfield  {journal} {\bibinfo  {journal} {Nat. Phys.}\ }\textbf {\bibinfo {volume} {14}},\ \bibinfo {pages} {495} (\bibinfo {year} {2018})}\BibitemShut {NoStop}%
\bibitem [{\citenamefont {Dyrda{\l}}\ \emph {et~al.}(2020)\citenamefont {Dyrda{\l}}, \citenamefont {Barna{\ifmmode\acute{s}\else\'{s}\fi}},\ and\ \citenamefont {Fert}}]{Dyrdal2020Jan}%
  \BibitemOpen
  \bibfield  {author} {\bibinfo {author} {\bibfnamefont {A.}~\bibnamefont {Dyrda{\l}}}, \bibinfo {author} {\bibfnamefont {J.}~\bibnamefont {Barna{\ifmmode\acute{s}\else\'{s}\fi}}},\ and\ \bibinfo {author} {\bibfnamefont {A.}~\bibnamefont {Fert}},\ }\bibfield  {title} {\bibinfo {title} {{Spin-Momentum-Locking Inhomogeneities as a Source of Bilinear Magnetoresistance in Topological Insulators}},\ }\href {https://doi.org/10.1103/PhysRevLett.124.046802} {\bibfield  {journal} {\bibinfo  {journal} {Phys. Rev. Lett.}\ }\textbf {\bibinfo {volume} {124}},\ \bibinfo {pages} {046802} (\bibinfo {year} {2020})}\BibitemShut {NoStop}%
\bibitem [{\citenamefont {Fu}\ \emph {et~al.}(2022)\citenamefont {Fu}, \citenamefont {Li}, \citenamefont {Papin}, \citenamefont {No{\ifmmode\ddot{e}\else\"{e}\fi}l}, \citenamefont {Teresi}, \citenamefont {Cosset-Ch{\ifmmode\acute{e}\else\'{e}\fi}neau}, \citenamefont {Grezes}, \citenamefont {Guillet}, \citenamefont {Thomas}, \citenamefont {Niquet}, \citenamefont {Ballet}, \citenamefont {Meunier}, \citenamefont {Attan{\ifmmode\acute{e}\else\'{e}\fi}}, \citenamefont {Fert},\ and\ \citenamefont {Vila}}]{Fu2022Oct}%
  \BibitemOpen
  \bibfield  {author} {\bibinfo {author} {\bibfnamefont {Y.}~\bibnamefont {Fu}}, \bibinfo {author} {\bibfnamefont {J.}~\bibnamefont {Li}}, \bibinfo {author} {\bibfnamefont {J.}~\bibnamefont {Papin}}, \bibinfo {author} {\bibfnamefont {P.}~\bibnamefont {No{\ifmmode\ddot{e}\else\"{e}\fi}l}}, \bibinfo {author} {\bibfnamefont {S.}~\bibnamefont {Teresi}}, \bibinfo {author} {\bibfnamefont {M.}~\bibnamefont {Cosset-Ch{\ifmmode\acute{e}\else\'{e}\fi}neau}}, \bibinfo {author} {\bibfnamefont {C.}~\bibnamefont {Grezes}}, \bibinfo {author} {\bibfnamefont {T.}~\bibnamefont {Guillet}}, \bibinfo {author} {\bibfnamefont {C.}~\bibnamefont {Thomas}}, \bibinfo {author} {\bibfnamefont {Y.-M.}\ \bibnamefont {Niquet}}, \bibinfo {author} {\bibfnamefont {P.}~\bibnamefont {Ballet}}, \bibinfo {author} {\bibfnamefont {T.}~\bibnamefont {Meunier}}, \bibinfo {author} {\bibfnamefont {J.-P.}\ \bibnamefont {Attan{\ifmmode\acute{e}\else\'{e}\fi}}}, \bibinfo {author} {\bibfnamefont {A.}~\bibnamefont {Fert}},\ and\ \bibinfo {author}
  {\bibfnamefont {L.}~\bibnamefont {Vila}},\ }\bibfield  {title} {\bibinfo {title} {{Bilinear Magnetoresistance in HgTe Topological Insulator: Opposite Signs at Opposite Surfaces Demonstrated by Gate Control}},\ }\href {https://doi.org/10.1021/acs.nanolett.2c02585} {\bibfield  {journal} {\bibinfo  {journal} {Nano Lett.}\ }\textbf {\bibinfo {volume} {22}},\ \bibinfo {pages} {7867} (\bibinfo {year} {2022})}\BibitemShut {NoStop}%
\bibitem [{\citenamefont {Guillet}\ \emph {et~al.}(2020)\citenamefont {Guillet}, \citenamefont {Zucchetti}, \citenamefont {Barbedienne}, \citenamefont {Marty}, \citenamefont {Isella}, \citenamefont {Cagnon}, \citenamefont {Vergnaud}, \citenamefont {Jaffr{\ifmmode\grave{e}\else\`{e}\fi}s}, \citenamefont {Reyren}, \citenamefont {George}, \citenamefont {Fert},\ and\ \citenamefont {Jamet}}]{Guillet2020Jan}%
  \BibitemOpen
  \bibfield  {author} {\bibinfo {author} {\bibfnamefont {T.}~\bibnamefont {Guillet}}, \bibinfo {author} {\bibfnamefont {C.}~\bibnamefont {Zucchetti}}, \bibinfo {author} {\bibfnamefont {Q.}~\bibnamefont {Barbedienne}}, \bibinfo {author} {\bibfnamefont {A.}~\bibnamefont {Marty}}, \bibinfo {author} {\bibfnamefont {G.}~\bibnamefont {Isella}}, \bibinfo {author} {\bibfnamefont {L.}~\bibnamefont {Cagnon}}, \bibinfo {author} {\bibfnamefont {C.}~\bibnamefont {Vergnaud}}, \bibinfo {author} {\bibfnamefont {H.}~\bibnamefont {Jaffr{\ifmmode\grave{e}\else\`{e}\fi}s}}, \bibinfo {author} {\bibfnamefont {N.}~\bibnamefont {Reyren}}, \bibinfo {author} {\bibfnamefont {J.-M.}\ \bibnamefont {George}}, \bibinfo {author} {\bibfnamefont {A.}~\bibnamefont {Fert}},\ and\ \bibinfo {author} {\bibfnamefont {M.}~\bibnamefont {Jamet}},\ }\bibfield  {title} {\bibinfo {title} {{Observation of Large Unidirectional Rashba Magnetoresistance in Ge(111)}},\ }\href {https://doi.org/10.1103/PhysRevLett.124.027201} {\bibfield  {journal} {\bibinfo
  {journal} {Phys. Rev. Lett.}\ }\textbf {\bibinfo {volume} {124}},\ \bibinfo {pages} {027201} (\bibinfo {year} {2020})}\BibitemShut {NoStop}%
\bibitem [{\citenamefont {Vaz}\ \emph {et~al.}(2020)\citenamefont {Vaz}, \citenamefont {Trier}, \citenamefont {Dyrda{\l}}, \citenamefont {Johansson}, \citenamefont {Garcia}, \citenamefont {Barth{\ifmmode\acute{e}\else\'{e}\fi}l{\ifmmode\acute{e}\else\'{e}\fi}my}, \citenamefont {Mertig}, \citenamefont {Barna{\ifmmode\acute{s}\else\'{s}\fi}}, \citenamefont {Fert},\ and\ \citenamefont {Bibes}}]{Vaz2020Jul}%
  \BibitemOpen
  \bibfield  {author} {\bibinfo {author} {\bibfnamefont {D.~C.}\ \bibnamefont {Vaz}}, \bibinfo {author} {\bibfnamefont {F.}~\bibnamefont {Trier}}, \bibinfo {author} {\bibfnamefont {A.}~\bibnamefont {Dyrda{\l}}}, \bibinfo {author} {\bibfnamefont {A.}~\bibnamefont {Johansson}}, \bibinfo {author} {\bibfnamefont {K.}~\bibnamefont {Garcia}}, \bibinfo {author} {\bibfnamefont {A.}~\bibnamefont {Barth{\ifmmode\acute{e}\else\'{e}\fi}l{\ifmmode\acute{e}\else\'{e}\fi}my}}, \bibinfo {author} {\bibfnamefont {I.}~\bibnamefont {Mertig}}, \bibinfo {author} {\bibfnamefont {J.}~\bibnamefont {Barna{\ifmmode\acute{s}\else\'{s}\fi}}}, \bibinfo {author} {\bibfnamefont {A.}~\bibnamefont {Fert}},\ and\ \bibinfo {author} {\bibfnamefont {M.}~\bibnamefont {Bibes}},\ }\bibfield  {title} {\bibinfo {title} {{Determining the Rashba parameter from the bilinear magnetoresistance response in a two-dimensional electron gas}},\ }\href {https://doi.org/10.1103/PhysRevMaterials.4.071001} {\bibfield  {journal} {\bibinfo  {journal} {Phys. Rev.
  Mater.}\ }\textbf {\bibinfo {volume} {4}},\ \bibinfo {pages} {071001} (\bibinfo {year} {2020})}\BibitemShut {NoStop}%
\bibitem [{\citenamefont {Sodemann}\ and\ \citenamefont {Fu}(2015)}]{Sodemann2015Nov}%
  \BibitemOpen
  \bibfield  {author} {\bibinfo {author} {\bibfnamefont {I.}~\bibnamefont {Sodemann}}\ and\ \bibinfo {author} {\bibfnamefont {L.}~\bibnamefont {Fu}},\ }\bibfield  {title} {\bibinfo {title} {{Quantum Nonlinear Hall Effect Induced by Berry Curvature Dipole in Time-Reversal Invariant Materials}},\ }\href {https://doi.org/10.1103/PhysRevLett.115.216806} {\bibfield  {journal} {\bibinfo  {journal} {Phys. Rev. Lett.}\ }\textbf {\bibinfo {volume} {115}},\ \bibinfo {pages} {216806} (\bibinfo {year} {2015})}\BibitemShut {NoStop}%
\bibitem [{\citenamefont {Ma}\ \emph {et~al.}(2019)\citenamefont {Ma}, \citenamefont {Xu}, \citenamefont {Shen}, \citenamefont {MacNeill}, \citenamefont {Fatemi}, \citenamefont {Chang}, \citenamefont {Mier~Valdivia}, \citenamefont {Wu}, \citenamefont {Du}, \citenamefont {Hsu}, \citenamefont {Fang}, \citenamefont {Gibson}, \citenamefont {Watanabe}, \citenamefont {Taniguchi}, \citenamefont {Cava}, \citenamefont {Kaxiras}, \citenamefont {Lu}, \citenamefont {Lin}, \citenamefont {Fu}, \citenamefont {Gedik},\ and\ \citenamefont {Jarillo-Herrero}}]{Ma2019Jan}%
  \BibitemOpen
  \bibfield  {author} {\bibinfo {author} {\bibfnamefont {Q.}~\bibnamefont {Ma}}, \bibinfo {author} {\bibfnamefont {S.-Y.}\ \bibnamefont {Xu}}, \bibinfo {author} {\bibfnamefont {H.}~\bibnamefont {Shen}}, \bibinfo {author} {\bibfnamefont {D.}~\bibnamefont {MacNeill}}, \bibinfo {author} {\bibfnamefont {V.}~\bibnamefont {Fatemi}}, \bibinfo {author} {\bibfnamefont {T.-R.}\ \bibnamefont {Chang}}, \bibinfo {author} {\bibfnamefont {A.~M.}\ \bibnamefont {Mier~Valdivia}}, \bibinfo {author} {\bibfnamefont {S.}~\bibnamefont {Wu}}, \bibinfo {author} {\bibfnamefont {Z.}~\bibnamefont {Du}}, \bibinfo {author} {\bibfnamefont {C.-H.}\ \bibnamefont {Hsu}}, \bibinfo {author} {\bibfnamefont {S.}~\bibnamefont {Fang}}, \bibinfo {author} {\bibfnamefont {Q.~D.}\ \bibnamefont {Gibson}}, \bibinfo {author} {\bibfnamefont {K.}~\bibnamefont {Watanabe}}, \bibinfo {author} {\bibfnamefont {T.}~\bibnamefont {Taniguchi}}, \bibinfo {author} {\bibfnamefont {R.~J.}\ \bibnamefont {Cava}}, \bibinfo {author} {\bibfnamefont {E.}~\bibnamefont
  {Kaxiras}}, \bibinfo {author} {\bibfnamefont {H.-Z.}\ \bibnamefont {Lu}}, \bibinfo {author} {\bibfnamefont {H.}~\bibnamefont {Lin}}, \bibinfo {author} {\bibfnamefont {L.}~\bibnamefont {Fu}}, \bibinfo {author} {\bibfnamefont {N.}~\bibnamefont {Gedik}},\ and\ \bibinfo {author} {\bibfnamefont {P.}~\bibnamefont {Jarillo-Herrero}},\ }\bibfield  {title} {\bibinfo {title} {{Observation of the nonlinear Hall effect under time-reversal-symmetric conditions}},\ }\href {https://doi.org/10.1038/s41586-018-0807-6} {\bibfield  {journal} {\bibinfo  {journal} {Nature}\ }\textbf {\bibinfo {volume} {565}},\ \bibinfo {pages} {337} (\bibinfo {year} {2019})}\BibitemShut {NoStop}%
\bibitem [{\citenamefont {Kang}\ \emph {et~al.}(2019)\citenamefont {Kang}, \citenamefont {Li}, \citenamefont {Sohn}, \citenamefont {Shan},\ and\ \citenamefont {Mak}}]{Kang2019Apr}%
  \BibitemOpen
  \bibfield  {author} {\bibinfo {author} {\bibfnamefont {K.}~\bibnamefont {Kang}}, \bibinfo {author} {\bibfnamefont {T.}~\bibnamefont {Li}}, \bibinfo {author} {\bibfnamefont {E.}~\bibnamefont {Sohn}}, \bibinfo {author} {\bibfnamefont {J.}~\bibnamefont {Shan}},\ and\ \bibinfo {author} {\bibfnamefont {K.~F.}\ \bibnamefont {Mak}},\ }\bibfield  {title} {\bibinfo {title} {{Nonlinear anomalous Hall effect in few-layer WTe2}},\ }\href {https://doi.org/10.1038/s41563-019-0294-7} {\bibfield  {journal} {\bibinfo  {journal} {Nat. Mater.}\ }\textbf {\bibinfo {volume} {18}},\ \bibinfo {pages} {324} (\bibinfo {year} {2019})}\BibitemShut {NoStop}%
\bibitem [{\citenamefont {He}\ \emph {et~al.}(2019)\citenamefont {He}, \citenamefont {Zhang}, \citenamefont {Zhu}, \citenamefont {Shi}, \citenamefont {Heinonen}, \citenamefont {Vignale},\ and\ \citenamefont {Yang}}]{He2019Jul}%
  \BibitemOpen
  \bibfield  {author} {\bibinfo {author} {\bibfnamefont {P.}~\bibnamefont {He}}, \bibinfo {author} {\bibfnamefont {S.~S.-L.}\ \bibnamefont {Zhang}}, \bibinfo {author} {\bibfnamefont {D.}~\bibnamefont {Zhu}}, \bibinfo {author} {\bibfnamefont {S.}~\bibnamefont {Shi}}, \bibinfo {author} {\bibfnamefont {O.~G.}\ \bibnamefont {Heinonen}}, \bibinfo {author} {\bibfnamefont {G.}~\bibnamefont {Vignale}},\ and\ \bibinfo {author} {\bibfnamefont {H.}~\bibnamefont {Yang}},\ }\bibfield  {title} {\bibinfo {title} {{Nonlinear Planar Hall Effect}},\ }\href {https://doi.org/10.1103/PhysRevLett.123.016801} {\bibfield  {journal} {\bibinfo  {journal} {Phys. Rev. Lett.}\ }\textbf {\bibinfo {volume} {123}},\ \bibinfo {pages} {016801} (\bibinfo {year} {2019})}\BibitemShut {NoStop}%
\bibitem [{\citenamefont {Du}\ \emph {et~al.}(2021)\citenamefont {Du}, \citenamefont {Wang}, \citenamefont {Sun}, \citenamefont {Lu},\ and\ \citenamefont {Xie}}]{Du2021Aug}%
  \BibitemOpen
  \bibfield  {author} {\bibinfo {author} {\bibfnamefont {Z.~Z.}\ \bibnamefont {Du}}, \bibinfo {author} {\bibfnamefont {C.~M.}\ \bibnamefont {Wang}}, \bibinfo {author} {\bibfnamefont {H.-P.}\ \bibnamefont {Sun}}, \bibinfo {author} {\bibfnamefont {H.-Z.}\ \bibnamefont {Lu}},\ and\ \bibinfo {author} {\bibfnamefont {X.~C.}\ \bibnamefont {Xie}},\ }\bibfield  {title} {\bibinfo {title} {{Quantum theory of the nonlinear Hall effect}},\ }\href {https://doi.org/10.1038/s41467-021-25273-4} {\bibfield  {journal} {\bibinfo  {journal} {Nat. Commun.}\ }\textbf {\bibinfo {volume} {12}},\ \bibinfo {pages} {1} (\bibinfo {year} {2021})}\BibitemShut {NoStop}%
\bibitem [{\citenamefont {Nagaosa}\ \emph {et~al.}(2010)\citenamefont {Nagaosa}, \citenamefont {Sinova}, \citenamefont {Onoda}, \citenamefont {MacDonald},\ and\ \citenamefont {Ong}}]{Sinova_RevModPhys_AHE_2010}%
  \BibitemOpen
  \bibfield  {author} {\bibinfo {author} {\bibfnamefont {N.}~\bibnamefont {Nagaosa}}, \bibinfo {author} {\bibfnamefont {J.}~\bibnamefont {Sinova}}, \bibinfo {author} {\bibfnamefont {S.}~\bibnamefont {Onoda}}, \bibinfo {author} {\bibfnamefont {A.~H.}\ \bibnamefont {MacDonald}},\ and\ \bibinfo {author} {\bibfnamefont {N.~P.}\ \bibnamefont {Ong}},\ }\bibfield  {title} {\bibinfo {title} {Anomalous hall effect},\ }\href {https://doi.org/10.1103/RevModPhys.82.1539} {\bibfield  {journal} {\bibinfo  {journal} {Rev. Mod. Phys.}\ }\textbf {\bibinfo {volume} {82}},\ \bibinfo {pages} {1539} (\bibinfo {year} {2010})}\BibitemShut {NoStop}%
\bibitem [{\citenamefont {Du}\ \emph {et~al.}(2019)\citenamefont {Du}, \citenamefont {Wang}, \citenamefont {Li}, \citenamefont {Lu},\ and\ \citenamefont {Xie}}]{Du2019Jul}%
  \BibitemOpen
  \bibfield  {author} {\bibinfo {author} {\bibfnamefont {Z.~Z.}\ \bibnamefont {Du}}, \bibinfo {author} {\bibfnamefont {C.~M.}\ \bibnamefont {Wang}}, \bibinfo {author} {\bibfnamefont {S.}~\bibnamefont {Li}}, \bibinfo {author} {\bibfnamefont {H.-Z.}\ \bibnamefont {Lu}},\ and\ \bibinfo {author} {\bibfnamefont {X.~C.}\ \bibnamefont {Xie}},\ }\bibfield  {title} {\bibinfo {title} {{Disorder-induced nonlinear Hall effect with time-reversal symmetry}},\ }\href {https://doi.org/10.1038/s41467-019-10941-3} {\bibfield  {journal} {\bibinfo  {journal} {Nat. Commun.}\ }\textbf {\bibinfo {volume} {10}},\ \bibinfo {pages} {1} (\bibinfo {year} {2019})}\BibitemShut {NoStop}%
\bibitem [{\citenamefont {Zhang}\ \emph {et~al.}(2018)\citenamefont {Zhang}, \citenamefont {Sun},\ and\ \citenamefont {Yan}}]{Zhang2018Jan}%
  \BibitemOpen
  \bibfield  {author} {\bibinfo {author} {\bibfnamefont {Y.}~\bibnamefont {Zhang}}, \bibinfo {author} {\bibfnamefont {Y.}~\bibnamefont {Sun}},\ and\ \bibinfo {author} {\bibfnamefont {B.}~\bibnamefont {Yan}},\ }\bibfield  {title} {\bibinfo {title} {{Berry curvature dipole in Weyl semimetal materials: An ab initio study}},\ }\href {https://doi.org/10.1103/PhysRevB.97.041101} {\bibfield  {journal} {\bibinfo  {journal} {Phys. Rev. B}\ }\textbf {\bibinfo {volume} {97}},\ \bibinfo {pages} {041101} (\bibinfo {year} {2018})}\BibitemShut {NoStop}%
\bibitem [{\citenamefont {You}\ \emph {et~al.}(2018)\citenamefont {You}, \citenamefont {Fang}, \citenamefont {Xu}, \citenamefont {Kaxiras},\ and\ \citenamefont {Low}}]{You2018Sep}%
  \BibitemOpen
  \bibfield  {author} {\bibinfo {author} {\bibfnamefont {J.-S.}\ \bibnamefont {You}}, \bibinfo {author} {\bibfnamefont {S.}~\bibnamefont {Fang}}, \bibinfo {author} {\bibfnamefont {S.-Y.}\ \bibnamefont {Xu}}, \bibinfo {author} {\bibfnamefont {E.}~\bibnamefont {Kaxiras}},\ and\ \bibinfo {author} {\bibfnamefont {T.}~\bibnamefont {Low}},\ }\bibfield  {title} {\bibinfo {title} {{Berry curvature dipole current in the transition metal dichalcogenides family}},\ }\href {https://doi.org/10.1103/PhysRevB.98.121109} {\bibfield  {journal} {\bibinfo  {journal} {Phys. Rev. B}\ }\textbf {\bibinfo {volume} {98}},\ \bibinfo {pages} {121109} (\bibinfo {year} {2018})}\BibitemShut {NoStop}%
\bibitem [{\citenamefont {Zhou}\ \emph {et~al.}(2020)\citenamefont {Zhou}, \citenamefont {Zhang},\ and\ \citenamefont {Law}}]{Zhou2020Feb}%
  \BibitemOpen
  \bibfield  {author} {\bibinfo {author} {\bibfnamefont {B.~T.}\ \bibnamefont {Zhou}}, \bibinfo {author} {\bibfnamefont {C.-P.}\ \bibnamefont {Zhang}},\ and\ \bibinfo {author} {\bibfnamefont {K.~T.}\ \bibnamefont {Law}},\ }\bibfield  {title} {\bibinfo {title} {{Highly Tunable Nonlinear Hall Effects Induced by Spin-Orbit Couplings in Strained Polar Transition-Metal Dichalcogenides}},\ }\href {https://doi.org/10.1103/PhysRevApplied.13.024053} {\bibfield  {journal} {\bibinfo  {journal} {Phys. Rev. Appl.}\ }\textbf {\bibinfo {volume} {13}},\ \bibinfo {pages} {024053} (\bibinfo {year} {2020})}\BibitemShut {NoStop}%
\bibitem [{\citenamefont {Nandy}\ and\ \citenamefont {Sodemann}(2019)}]{Nandy2019Nov}%
  \BibitemOpen
  \bibfield  {author} {\bibinfo {author} {\bibfnamefont {S.}~\bibnamefont {Nandy}}\ and\ \bibinfo {author} {\bibfnamefont {I.}~\bibnamefont {Sodemann}},\ }\bibfield  {title} {\bibinfo {title} {{Symmetry and quantum kinetics of the nonlinear Hall effect}},\ }\href {https://doi.org/10.1103/PhysRevB.100.195117} {\bibfield  {journal} {\bibinfo  {journal} {Phys. Rev. B}\ }\textbf {\bibinfo {volume} {100}},\ \bibinfo {pages} {195117} (\bibinfo {year} {2019})}\BibitemShut {NoStop}%
\bibitem [{\citenamefont {Bernevig}\ \emph {et~al.}(2006)\citenamefont {Bernevig}, \citenamefont {Hughes},\ and\ \citenamefont {Zhang}}]{Bernevig2006Dec}%
  \BibitemOpen
  \bibfield  {author} {\bibinfo {author} {\bibfnamefont {B.~A.}\ \bibnamefont {Bernevig}}, \bibinfo {author} {\bibfnamefont {T.~L.}\ \bibnamefont {Hughes}},\ and\ \bibinfo {author} {\bibfnamefont {S.-C.}\ \bibnamefont {Zhang}},\ }\bibfield  {title} {\bibinfo {title} {{Quantum Spin Hall Effect and Topological Phase Transition in HgTe Quantum Wells}},\ }\href {https://doi.org/10.1126/science.1133734} {\bibfield  {journal} {\bibinfo  {journal} {Science}\ }\textbf {\bibinfo {volume} {314}},\ \bibinfo {pages} {1757} (\bibinfo {year} {2006})}\BibitemShut {NoStop}%
\bibitem [{\citenamefont {Schliemann}\ and\ \citenamefont {Loss}(2005)}]{Schliemann2005Feb}%
  \BibitemOpen
  \bibfield  {author} {\bibinfo {author} {\bibfnamefont {J.}~\bibnamefont {Schliemann}}\ and\ \bibinfo {author} {\bibfnamefont {D.}~\bibnamefont {Loss}},\ }\bibfield  {title} {\bibinfo {title} {{Spin-Hall transport of heavy holes in III-V semiconductor quantum wells}},\ }\href {https://doi.org/10.1103/PhysRevB.71.085308} {\bibfield  {journal} {\bibinfo  {journal} {Phys. Rev. B}\ }\textbf {\bibinfo {volume} {71}},\ \bibinfo {pages} {085308} (\bibinfo {year} {2005})}\BibitemShut {NoStop}%
\bibitem [{\citenamefont {Liu}\ \emph {et~al.}(2008)\citenamefont {Liu}, \citenamefont {Zhou}, \citenamefont {Shen},\ and\ \citenamefont {Zhu}}]{Liu2008Mar}%
  \BibitemOpen
  \bibfield  {author} {\bibinfo {author} {\bibfnamefont {C.-X.}\ \bibnamefont {Liu}}, \bibinfo {author} {\bibfnamefont {B.}~\bibnamefont {Zhou}}, \bibinfo {author} {\bibfnamefont {S.-Q.}\ \bibnamefont {Shen}},\ and\ \bibinfo {author} {\bibfnamefont {B.-f.}\ \bibnamefont {Zhu}},\ }\bibfield  {title} {\bibinfo {title} {{Current-induced spin polarization in a two-dimensional hole gas}},\ }\href {https://doi.org/10.1103/PhysRevB.77.125345} {\bibfield  {journal} {\bibinfo  {journal} {Phys. Rev. B}\ }\textbf {\bibinfo {volume} {77}},\ \bibinfo {pages} {125345} (\bibinfo {year} {2008})}\BibitemShut {NoStop}%
\bibitem [{\citenamefont {Moriya}\ \emph {et~al.}(2014)\citenamefont {Moriya}, \citenamefont {Sawano}, \citenamefont {Hoshi}, \citenamefont {Masubuchi}, \citenamefont {Shiraki}, \citenamefont {Wild}, \citenamefont {Neumann}, \citenamefont {Abstreiter}, \citenamefont {Bougeard}, \citenamefont {Koga},\ and\ \citenamefont {Machida}}]{Moriya2014Aug}%
  \BibitemOpen
  \bibfield  {author} {\bibinfo {author} {\bibfnamefont {R.}~\bibnamefont {Moriya}}, \bibinfo {author} {\bibfnamefont {K.}~\bibnamefont {Sawano}}, \bibinfo {author} {\bibfnamefont {Y.}~\bibnamefont {Hoshi}}, \bibinfo {author} {\bibfnamefont {S.}~\bibnamefont {Masubuchi}}, \bibinfo {author} {\bibfnamefont {Y.}~\bibnamefont {Shiraki}}, \bibinfo {author} {\bibfnamefont {A.}~\bibnamefont {Wild}}, \bibinfo {author} {\bibfnamefont {C.}~\bibnamefont {Neumann}}, \bibinfo {author} {\bibfnamefont {G.}~\bibnamefont {Abstreiter}}, \bibinfo {author} {\bibfnamefont {D.}~\bibnamefont {Bougeard}}, \bibinfo {author} {\bibfnamefont {T.}~\bibnamefont {Koga}},\ and\ \bibinfo {author} {\bibfnamefont {T.}~\bibnamefont {Machida}},\ }\bibfield  {title} {\bibinfo {title} {{Cubic Rashba Spin-Orbit Interaction of a Two-Dimensional Hole Gas in a Strained-$\mathrm{Ge}/\mathrm{SiGe}$ Quantum Well}},\ }\href {https://doi.org/10.1103/PhysRevLett.113.086601} {\bibfield  {journal} {\bibinfo  {journal} {Phys. Rev. Lett.}\ }\textbf {\bibinfo
  {volume} {113}},\ \bibinfo {pages} {086601} (\bibinfo {year} {2014})}\BibitemShut {NoStop}%
\bibitem [{\citenamefont {van Heeringen}\ \emph {et~al.}(2017)\citenamefont {van Heeringen}, \citenamefont {McCollam}, \citenamefont {de~Wijs},\ and\ \citenamefont {Fasolino}}]{vanHeeringen2017Apr}%
  \BibitemOpen
  \bibfield  {author} {\bibinfo {author} {\bibfnamefont {L.~W.}\ \bibnamefont {van Heeringen}}, \bibinfo {author} {\bibfnamefont {A.}~\bibnamefont {McCollam}}, \bibinfo {author} {\bibfnamefont {G.~A.}\ \bibnamefont {de~Wijs}},\ and\ \bibinfo {author} {\bibfnamefont {A.}~\bibnamefont {Fasolino}},\ }\bibfield  {title} {\bibinfo {title} {{Theoretical models of Rashba spin splitting in asymmetric ${\mathrm{SrTiO}}_{3}$-based heterostructures}},\ }\href {https://doi.org/10.1103/PhysRevB.95.155134} {\bibfield  {journal} {\bibinfo  {journal} {Phys. Rev. B}\ }\textbf {\bibinfo {volume} {95}},\ \bibinfo {pages} {155134} (\bibinfo {year} {2017})}\BibitemShut {NoStop}%
\bibitem [{\citenamefont {Nakamura}\ \emph {et~al.}(2012)\citenamefont {Nakamura}, \citenamefont {Koga},\ and\ \citenamefont {Kimura}}]{Nakamura2012May}%
  \BibitemOpen
  \bibfield  {author} {\bibinfo {author} {\bibfnamefont {H.}~\bibnamefont {Nakamura}}, \bibinfo {author} {\bibfnamefont {T.}~\bibnamefont {Koga}},\ and\ \bibinfo {author} {\bibfnamefont {T.}~\bibnamefont {Kimura}},\ }\bibfield  {title} {\bibinfo {title} {{Experimental Evidence of Cubic Rashba Effect in an Inversion-Symmetric Oxide}},\ }\href {https://doi.org/10.1103/PhysRevLett.108.206601} {\bibfield  {journal} {\bibinfo  {journal} {Phys. Rev. Lett.}\ }\textbf {\bibinfo {volume} {108}},\ \bibinfo {pages} {206601} (\bibinfo {year} {2012})}\BibitemShut {NoStop}%
\bibitem [{\citenamefont {Liang}\ \emph {et~al.}(2015)\citenamefont {Liang}, \citenamefont {Cheng}, \citenamefont {Wei}, \citenamefont {Luo}, \citenamefont {Yu}, \citenamefont {Zeng},\ and\ \citenamefont {Zhang}}]{Liang2015Aug}%
  \BibitemOpen
  \bibfield  {author} {\bibinfo {author} {\bibfnamefont {H.}~\bibnamefont {Liang}}, \bibinfo {author} {\bibfnamefont {L.}~\bibnamefont {Cheng}}, \bibinfo {author} {\bibfnamefont {L.}~\bibnamefont {Wei}}, \bibinfo {author} {\bibfnamefont {Z.}~\bibnamefont {Luo}}, \bibinfo {author} {\bibfnamefont {G.}~\bibnamefont {Yu}}, \bibinfo {author} {\bibfnamefont {C.}~\bibnamefont {Zeng}},\ and\ \bibinfo {author} {\bibfnamefont {Z.}~\bibnamefont {Zhang}},\ }\bibfield  {title} {\bibinfo {title} {{Nonmonotonically tunable Rashba spin-orbit coupling by multiple-band filling control in ${\mathrm{SrTiO}}_{3}$-based interfacial $d$-electron gases}},\ }\href {https://doi.org/10.1103/PhysRevB.92.075309} {\bibfield  {journal} {\bibinfo  {journal} {Phys. Rev. B}\ }\textbf {\bibinfo {volume} {92}},\ \bibinfo {pages} {075309} (\bibinfo {year} {2015})}\BibitemShut {NoStop}%
\bibitem [{\citenamefont {Caviglia}\ \emph {et~al.}(2010)\citenamefont {Caviglia}, \citenamefont {Gabay}, \citenamefont {Gariglio}, \citenamefont {Reyren}, \citenamefont {Cancellieri},\ and\ \citenamefont {Triscone}}]{Caviglia2010Mar}%
  \BibitemOpen
  \bibfield  {author} {\bibinfo {author} {\bibfnamefont {A.~D.}\ \bibnamefont {Caviglia}}, \bibinfo {author} {\bibfnamefont {M.}~\bibnamefont {Gabay}}, \bibinfo {author} {\bibfnamefont {S.}~\bibnamefont {Gariglio}}, \bibinfo {author} {\bibfnamefont {N.}~\bibnamefont {Reyren}}, \bibinfo {author} {\bibfnamefont {C.}~\bibnamefont {Cancellieri}},\ and\ \bibinfo {author} {\bibfnamefont {J.-M.}\ \bibnamefont {Triscone}},\ }\bibfield  {title} {\bibinfo {title} {{Tunable Rashba Spin-Orbit Interaction at Oxide Interfaces}},\ }\href {https://doi.org/10.1103/PhysRevLett.104.126803} {\bibfield  {journal} {\bibinfo  {journal} {Phys. Rev. Lett.}\ }\textbf {\bibinfo {volume} {104}},\ \bibinfo {pages} {126803} (\bibinfo {year} {2010})}\BibitemShut {NoStop}%
\bibitem [{\citenamefont {Lesne}\ \emph {et~al.}(2016)\citenamefont {Lesne}, \citenamefont {Fu}, \citenamefont {Oyarzun}, \citenamefont {Rojas-S{\ifmmode\acute{a}\else\'{a}\fi}nchez}, \citenamefont {Vaz}, \citenamefont {Naganuma}, \citenamefont {Sicoli}, \citenamefont {Attan{\ifmmode\acute{e}\else\'{e}\fi}}, \citenamefont {Jamet}, \citenamefont {Jacquet}, \citenamefont {George}, \citenamefont {Barth{\ifmmode\acute{e}\else\'{e}\fi}l{\ifmmode\acute{e}\else\'{e}\fi}my}, \citenamefont {Jaffr{\ifmmode\grave{e}\else\`{e}\fi}s}, \citenamefont {Fert}, \citenamefont {Bibes},\ and\ \citenamefont {Vila}}]{Lesne2016Dec}%
  \BibitemOpen
  \bibfield  {author} {\bibinfo {author} {\bibfnamefont {E.}~\bibnamefont {Lesne}}, \bibinfo {author} {\bibfnamefont {Y.}~\bibnamefont {Fu}}, \bibinfo {author} {\bibfnamefont {S.}~\bibnamefont {Oyarzun}}, \bibinfo {author} {\bibfnamefont {J.~C.}\ \bibnamefont {Rojas-S{\ifmmode\acute{a}\else\'{a}\fi}nchez}}, \bibinfo {author} {\bibfnamefont {D.~C.}\ \bibnamefont {Vaz}}, \bibinfo {author} {\bibfnamefont {H.}~\bibnamefont {Naganuma}}, \bibinfo {author} {\bibfnamefont {G.}~\bibnamefont {Sicoli}}, \bibinfo {author} {\bibfnamefont {J.-P.}\ \bibnamefont {Attan{\ifmmode\acute{e}\else\'{e}\fi}}}, \bibinfo {author} {\bibfnamefont {M.}~\bibnamefont {Jamet}}, \bibinfo {author} {\bibfnamefont {E.}~\bibnamefont {Jacquet}}, \bibinfo {author} {\bibfnamefont {J.-M.}\ \bibnamefont {George}}, \bibinfo {author} {\bibfnamefont {A.}~\bibnamefont {Barth{\ifmmode\acute{e}\else\'{e}\fi}l{\ifmmode\acute{e}\else\'{e}\fi}my}}, \bibinfo {author} {\bibfnamefont {H.}~\bibnamefont {Jaffr{\ifmmode\grave{e}\else\`{e}\fi}s}}, \bibinfo {author}
  {\bibfnamefont {A.}~\bibnamefont {Fert}}, \bibinfo {author} {\bibfnamefont {M.}~\bibnamefont {Bibes}},\ and\ \bibinfo {author} {\bibfnamefont {L.}~\bibnamefont {Vila}},\ }\bibfield  {title} {\bibinfo {title} {{Highly efficient and tunable spin-to-charge conversion through Rashba coupling at oxide interfaces}},\ }\href {https://doi.org/10.1038/nmat4726} {\bibfield  {journal} {\bibinfo  {journal} {Nat. Mater.}\ }\textbf {\bibinfo {volume} {15}},\ \bibinfo {pages} {1261} (\bibinfo {year} {2016})}\BibitemShut {NoStop}%
\bibitem [{\citenamefont {Trier}\ \emph {et~al.}(2022)\citenamefont {Trier}, \citenamefont {No{\ifmmode\ddot{e}\else\"{e}\fi}l}, \citenamefont {Kim}, \citenamefont {Attan{\ifmmode\acute{e}\else\'{e}\fi}}, \citenamefont {Vila},\ and\ \citenamefont {Bibes}}]{Trier2022Apr}%
  \BibitemOpen
  \bibfield  {author} {\bibinfo {author} {\bibfnamefont {F.}~\bibnamefont {Trier}}, \bibinfo {author} {\bibfnamefont {P.}~\bibnamefont {No{\ifmmode\ddot{e}\else\"{e}\fi}l}}, \bibinfo {author} {\bibfnamefont {J.-V.}\ \bibnamefont {Kim}}, \bibinfo {author} {\bibfnamefont {J.-P.}\ \bibnamefont {Attan{\ifmmode\acute{e}\else\'{e}\fi}}}, \bibinfo {author} {\bibfnamefont {L.}~\bibnamefont {Vila}},\ and\ \bibinfo {author} {\bibfnamefont {M.}~\bibnamefont {Bibes}},\ }\bibfield  {title} {\bibinfo {title} {{Oxide spin-orbitronics: spin{\textendash}charge interconversion and topological spin textures}},\ }\href {https://doi.org/10.1038/s41578-021-00395-9} {\bibfield  {journal} {\bibinfo  {journal} {Nat. Rev. Mater.}\ }\textbf {\bibinfo {volume} {7}},\ \bibinfo {pages} {258} (\bibinfo {year} {2022})}\BibitemShut {NoStop}%
\bibitem [{\citenamefont {Karwacki}\ \emph {et~al.}(2018)\citenamefont {Karwacki}, \citenamefont {Dyrda\l{}}, \citenamefont {Berakdar},\ and\ \citenamefont {Barna\ifmmode~\acute{s}\else \'{s}\fi{}}}]{Karwacki_PRB_2018}%
  \BibitemOpen
  \bibfield  {author} {\bibinfo {author} {\bibfnamefont {L.}~\bibnamefont {Karwacki}}, \bibinfo {author} {\bibfnamefont {A.}~\bibnamefont {Dyrda\l{}}}, \bibinfo {author} {\bibfnamefont {J.}~\bibnamefont {Berakdar}},\ and\ \bibinfo {author} {\bibfnamefont {J.}~\bibnamefont {Barna\ifmmode~\acute{s}\else \'{s}\fi{}}},\ }\bibfield  {title} {\bibinfo {title} {{Current-induced spin polarization in the isotropic $k$-cubed Rashba model: Theoretical study of $p$-doped semiconductor heterostructures and perovskite-oxide interfaces}},\ }\href {https://doi.org/10.1103/PhysRevB.97.235302} {\bibfield  {journal} {\bibinfo  {journal} {Phys. Rev. B}\ }\textbf {\bibinfo {volume} {97}},\ \bibinfo {pages} {235302} (\bibinfo {year} {2018})}\BibitemShut {NoStop}%
\bibitem [{Note1()}]{Note1}%
  \BibitemOpen
  \bibinfo {note} {We do not consider here the skew-scattering and side jump processes.}\BibitemShut {Stop}%
\bibitem [{\citenamefont {Krzyżewska}\ \emph {et~al.}()\citenamefont {Krzyżewska}, \citenamefont {Dyrdał}, \citenamefont {Barnaś},\ and\ \citenamefont {Berakdar}}]{Krzyzewska2018}%
  \BibitemOpen
  \bibfield  {author} {\bibinfo {author} {\bibfnamefont {A.}~\bibnamefont {Krzyżewska}}, \bibinfo {author} {\bibfnamefont {A.}~\bibnamefont {Dyrdał}}, \bibinfo {author} {\bibfnamefont {J.}~\bibnamefont {Barnaś}},\ and\ \bibinfo {author} {\bibfnamefont {J.}~\bibnamefont {Berakdar}},\ }\bibfield  {title} {\bibinfo {title} {{Anomalous Hall and Nernst Effects in 2D Systems: Role of Cubic Rashba Spin–Orbit Coupling}},\ }\href {https://doi.org/https://doi.org/10.1002/pssr.201800232} {\bibfield  {journal} {\bibinfo  {journal} {physica status solidi (RRL) – Rapid Research Letters}\ }\textbf {\bibinfo {volume} {12}},\ \bibinfo {pages} {1800232}}\BibitemShut {NoStop}%
\end{thebibliography}%

\end{document}